\documentstyle[12pt,epsf]{article}

\newcommand\putfig[3]{
   \vbox{
   \let\picnaturalsize=N
   \def\picsize{#3}
   \def\picfilename{#1}
   \ifx\nopictures Y\else{\ifx\epsfloaded Y\else\input epsf \fi
   \let\epsfloaded=Y
   \centerline{\ifx\picnaturalsize N\epsfxsize \picsize\fi
   \epsfbox{\picfilename}}}\fi
   \vspace{1.0cm}
   {\it #2}
   \vspace{1.5cm}
   }
}

\def\hlf{{1 \over 2}}

\def\a{{\alpha}}
\def\b{{\beta}}
\def\g{{\gamma}}
\def\L{{\Lambda}}
\def\eN{{{\epsilon\over N}}}
\def\hP{{\hat{P}}}
\def\bW{{W^{*}}}

\def\bra{{\langle}}
\def\ket{{\rangle}}
\def\lA{{\lambda A}}
\newcommand\dl[2]{{\delta_{#1}^{{}{#2}}}}
\newcommand\xP[1]{{\hP_{\{#1\}}}}

\def\tn{{\tilde{n}}}
\def\bp{{\bar{p}}}

\def\bD{{\bar{a}}}
\def\bT{{\bar{a}^\dagger}}
\def\D{{a}}
\def\T{{a^\dagger}}

\def\cG{{\cal G}}
\def\cF{{\cal F}}

\def\vN{{1 \over N}}
\def\vNN{{1 \over {N^2}}}
\newcommand\pbl[1]{{\bar{p}^{({#1})}}}
\newcommand\binom[2]{{{#1}\choose {#2}}}

\def\gYM{$gYM_2$\ }
\def\YM{$YM_2$\ }
\def\be{\begin{equation}}
\def\ee{\end{equation}}

\def\cmp#1{{\it Comm. Math. Phys.} {\bf #1}}

\def\pl#1{{\it Phys. Lett.} {\bf #1B}}

\def\prd#1{{\it Phys. Rev.} {\bf D#1}}

\def\np#1{{\it Nucl. Phys.} {\bf B#1}}

\def\jmath#1{{\it J. Math. Phys.} {\bf #1}}
\def\mpl#1{{\it Mod. Phys. Lett.}{\bf A#1}}

\begin{document}

\begin{titlepage}
\titlepage
\rightline{TAUP-2182-94}
\rightline{hep-th/yymmddd}
\rightline{\today}
\vskip 1cm
\centerline {{\Large \bf The String Theory Approach to}}
\centerline {{\Large \bf Generalized 2D Yang-Mills Theory}}
\vskip 1cm

\centerline {O. Ganor , J. Sonnenschein and S. Yankielowicz
\footnote{Work supported in part by the US-Israel Binational Science
Foundation, `GIF' -- the German-Israeli Foundation
for Scientific Research and Development and the Israel Academy of Sciences.}}

\vskip 1cm

\begin{center}

\em  School of Physics and Astronomy\\

Beverly and Raymond Sackler \\

Faculty of Exact Sciences\\

Tel-Aviv University\\

Ramat Aviv Tel-Aviv, 69987, Israel

\end{center}

\vskip 1cm

\abstract{
We calculate the partition function of the $SU(N)$ ( and $U(N)$)
 generalized $YM_2$ theory defined on an arbitrary  Riemann  surface. The
result
which is  expressed as a sum over irreducible representations  generalizes the
Rusakov formula for ordinary \YM theory.
A diagrammatic expansion of the formula enables us to derive a Gross-Taylor
like
stringy description of the  model.
A sum of  2D string maps is shown to
reproduce the gauge theory results.
Maps  with branch points of degree higher than one, as well as ``microscopic
surfaces'' play an important role in the sum.
We discuss the underlying string theory.
}

\end{titlepage}

\section{Introduction}
The stringy interpretation of four dimensional
$YM$ and $QCD$ theories  is one of the longstanding
problems of strong interactions. As a matter of fact, the string theory's
first incarnation came about as the theory of strongly interacting hadrons.
Strong coupling lattice calculations as well as the large $N$ approach
clearly support and hint toward a string theory representation of these
gauge theories. Recently, in a beautiful series of
papers\cite{GT,GT2,GT3,GT4} this
problem was investigated by Gross and Taylor within the framework of $YM_2$
theory on a Riemann surface of arbitrary topology.
A lattice version of 2D YM ($YM_2$) theory
has been known for a long time  to be  exactly solvable\cite{Mig}.
 Several   algorithms to compute
correlators of Wilson loops on the plane
were written down in this framework\cite{Kaza,Barlic}.
More recently the partition function on an arbitrary Riemann surface was
shown to have a simple expression in terms of a sum over all irreducible
representations of the gauge group \cite{Rus}.
 An identical result was derived also in a
continuum path-integral approach by employing   a ``perturbation" to  the
topological gauge theory \cite{WitRev}.

 By  performing
a large $N$ expansion  of these results
 Gross and Taylor\cite{GT,GT2,GT3,GT4} managed to give a string interpretation
to the theory. To be more precise, they have shown that the large $N$ expansion
can be thought of as a string expansion with $g_{st}={1\over N}$
and with the string tension identified with $g^2 N$, where $g$ is the
gauge coupling.
  The coefficients of this expansion have a geometrical meaning in terms
of sums over maps between two dimensional orientable
 manifolds. They were reproduced  by attributing
 specific weights to different classes of
singular maps \cite{Mina,GT3,GT4}.
 Maps from non-orientable world-sheets were later shown to be
related to the stringy nature of $SO(N)$ and $Sp(N)$ gauge
theories\cite{NRS,Ramg}.
 A stringy behaviour was revealed recently also
for \YM at finite $N$ \cite{BaTa}.

 One of the more challenging problems in this context
is to find a string action which
gives rise to the sums of maps that reproduce the gauge  theory
partition function.
Recently, an important progress toward this goal has been
made\cite{Horava,CMR}.
These works indicate that the action may very well be a topological string
action.  This is consistent with the observation, which was noted previously
\cite{BCOV}, that the topological string action associated with holomorphic
maps correctly accounts for the leading ${1\over N}$ behaviour of the
partition function on a toroidal target space.

 Pure \YM theory defined on a compact Riemann surface is characterized by its
 invariance under area preserving diffeomorphisms  and by the fact that
there are no propagating degrees of freedom.
It is easy to realize that these properties are not unique to the
$Tr(F^2)$ theory but rather are shared by a wide class of theories,
 the generalized Yang-Mills theories ($gYM_2$).
An equivalent formulation  of the \YM Lagrangian  takes the form  of
$Tr(BF) +g^2 Tr (B^2)$, where $B$ is an auxiliary pseudo-scalar  field in the
adjoint of the group. The generalized theories correspond to a general
function $\Phi(B)$ replacing the $Tr(B^2)$ term.
These theories  can be further coupled  to fermions, thus obtaining the
 generalized $QCD_2$ theory \cite{DougLi}.
The interest in generalized $YM_2$ and $QCD_2$ theories stems from
the following reasons:
\begin{itemize}
\item 
Within the framework of 2D gauge theories, they  can serve as  new toy models.
 As laboratory
models for $YM_4$ and $QCD_4$ there is a-priori no special role to ordinary
$YM_2$ and $QCD_2$. As remarked in \cite{DougLi}, it is conceivable
that one of the generalized 2D models will reveal features which are more
relevant and more closely resemble the four dimensional theories of
interest.
\item 
The investigation of  the \gYM  gives information about correlators
of the  topological $BF$ theory and the ordinary \YM theory.
\item 
The study of  the stringy generalized $YM_2$  sheds additional light
on the underlying  topological string theory. This is mainly due to the
 singular maps of higher order branch points  which
do not show up in the ordinary theory and play an important role in the
generalized one.
These maps reside on the boundary of map space.
\item 
In four dimensions, $YM_4$ and $QCD_4$ do share a unique position
as renormalizable theories. The generalized theories in four dimensions seem
to introduce non-renormalizable terms. Starting from the lattice version
of these theories, such terms, which do not affect the continuum IR behaviour,
can be important from a computational point of view. We refer the reader
to \cite{DougLi} for further discussions of this point.
\end{itemize}

In the present work we show that following  the same procedure
used in    the ordinary
$YM_2$ theory all the generalized
theories   are  exactly solvable.
The partition function of the $SU(N)$ ( and $U(N)$) models
 defined on an arbitrary  Riemann  surface is shown to be expressed by a
formula that is a straightforward generalization of the \YM
result \cite{Rus,WitGauge},
involving dimensions and higher order Casimir operators.
 We introduce a diagrammatic expansion for higher Casimir operators.
By  using the latter we  derive a Gross-Taylor like
stringy description of the  model. In this description
a sum of  2D string maps that
reproduces the  gauge theory results.
In terms of those maps the distinction between the various models  is done by
assigning different weights to
 branch points of degree higher than one.
A concept of contracted microscopic surfaces
generalizes the contracted handles of \cite{Mina,GT3,GT4}.
Some preliminary ideas about the action of the underlying topological string
theory are sketched in relation to the Lagrangian proposed in \cite{CMR}.

The paper is organized as follows:
In section (2) we review the derivation of the partition function of
the ordinary (quadratic) $YM_2$ theory. We then derive the analogous result for
the \gYM theory  and sketch rules for the corresponding computations of loop
averages.
Section (3) is devoted to the derivation of the stringy picture of the
generalized models.
We start with a brief summary of the work of \cite{GT3,GT4}.
We explain the necessity to express the $SU(N)$  higher Casimir operators  in
terms of properties of the symmetric group $S_n$ in order to establish
the stringy picture.
This is followed by a subsection where we develop  in great details
a diagrammatic expansion of the Casimir operators.
Readers who are not interested in the derivation of the diagrammatic expansion
may skip this subsection and proceed directly to the one where the
rules of the cycle structure formula are stated.  This then leads to the
description of the partition function as a sum of maps. We encourage the reader
to go over the various examples presented in subsection 3.6.
Marked points and collapsed surfaces which are among the new features of the
generalized model are then discussed.
In section 4 we summarize the results of the work, present preliminary
thoughts about the underlying topological string action and state certain
open questions.
  In appendix (A) we
show that the same scaling rules used to determine the stringy maps
are obtained from a t'Hooft like analysis
of a generalized $YM_4$ or a generalized $QCD_2$.
In appendix (B) we compare the stringy calculation of certain correlators
(e.g. $tr(F^3)$) in the \YM theory to the corresponding correlators
(e.g. $tr(B^3)$) derived by employing the \gYM results.

\section{Solution of the \gYM}
Pure \YM theory defined on an arbitrary Riemann surface is known to be exactly
solvable.  In one approach the theory was regularized on the lattice\cite{Mig}
 and, using a heat kernel action, explicit expressions for the partition
function and loop averages  were derived.
Identical results were derived also in a
continuum path-integral approach \cite{WitRev}. In the following subsections
 we briefly review the former derivation of the  partition function and
 determine in a similar way
the results for the partition function and Wilson loops  in the \gYM case.

 \subsection{ The partition function the \YM theory}
The partition function for the ordinary $YM_2$ theory defined on a
compact Riemann surface ${\cal M}$ of genus $H$ and area $A$ is
\begin{equation}
   {\cal Z}(N,H,\lambda A) = \int [DA^{\mu}]exp[-{{1}\over {4 g^2}}
\int_{\Sigma} d^2x \sqrt{\det G_{ij}} ~tr F_{ij}F^{ij}]  \nonumber\\
\end{equation}
where the gauge group $G$ is taken to be either $SU(N)$ or $U(N)$,
$g$ is the gauge coupling constant, $\lambda= g^2 N$,
$G_{ij}$ is the metric on ${\cal M}$,
and  $tr$  stands for the trace in the fundamental representation.
The lattice partition function defined on an arbitrary triangulation of the
surface  is given by
\begin{equation}
   {\cal Z}_{\cal M}= \int \prod_l dU_l \prod_\triangle
Z_\triangle[U_\triangle]
  \nonumber\\
\end{equation}
where $\prod_l$ denotes a product over all links,   $U_\triangle$ is the
holonomy around a plaquette, and $Z_\triangle$ is a plaquette action.
For the latter  one uses a heat kernel action \cite{Mig,Rus}
rather than the Wilson action, i.e.
\begin{equation}
 Z_\triangle[U] =  \sum_R d_R \chi_R(U) e^{-t c_2(R)}
\label{QUAD}
\end{equation}
where the summation is over the irreducible representations $R$ of the group.
$d_R, \  \chi_R(U)$ and   $c_2(R)$ denote the dimension, character of $U$ and
the second Cassimir operator of $R$ respectively, and   $t = g^2 a$ with $a^2$
being the plaquette area.
 The holonomy $U$ (its  subscript $\triangle$
is omitted from  here on)
behaves as $U\approx 1-ia F$ when $a$ is small.
 Note that the region of validity of (\ref{QUAD}) is not only
$a\rightarrow 0$ with $F$ fixed, but actually also
$a\rightarrow 0$ with $F$ going to infinity as $a^{-1/2}$ because
this is the region for which the exponential $-{1\over {4 t}} a^2 Tr F^2$
is of order unity.

We will briefly review the derivation which singles out (\ref{QUAD})
as a convenient choice among the different lattice theories which
belong to the same universality class \cite{Mig,Rus}.
Let us look for a function $\Psi(U,t)$ that will replace
the continuum $e^{-{1\over {4 t}} a^2 Tr F^2}$.

The requirements  which we impose on $\Psi$ are
\begin{enumerate}
\item
As $t$ goes to zero  (and, therefore, for finite $g$
 also $a$ goes to zero) we want the holonomy to be close
to $1$:
$$\Psi(U,0) = \delta(U-1)$$

\item
For any $V\in G$ we have
$$\Psi(V^{-1}UV, t) = \Psi(U,t)$$
In other words $\Psi$ is a {\em class-function}.

\item
$\Psi$ satisfies the heat kernel equation
$$
\{ {\partial\over {\partial t}} -\sum_{a,b}g_{ab}
\partial^a\partial^b \} \Psi(U,t) = 0
$$
where $g_{ab}$ is the inverse of the Cartan metric
$$
g^{ab} = tr (t^a t^b)
$$
\end{enumerate}

To see that (\ref{QUAD}) is an approximate solution to the heat-kernel equation
we note that any class function is a linear combination of characters.
The differentiation of a character in the direction of a Lie algebra
element $t^a$ is given by:
$$
\partial^{a_1}\partial^{a_2}\cdots\partial^{a_k}
\chi_R(U)
 = {{i^k}\over{k!}}\chi_R(U t^{(a_1} t^{a_2} \cdots t^{a_k)}) + O(U-1)
$$
The notation $\chi_R(U t^{a_1} t^{a_2} \cdots t^{a_k})$
stands for  the trace of the multiplication
of the matrices which represent $U$ and $t^{a_1},\dots,t^{a_k}$
in the representation $R$. The brackets $(\cdots)$ imply symmetrization
with respect to the indices.
The term  $O(U-1)$ means that the corrections are of the order of
$U-1 \sim aF\sim t^{1/2}$.
Since
$$
\sum_{a,b}g_{ab}\partial^a\partial^b \chi_R(U) \approx
-\hlf\chi_R(U \sum_{a,b}g_{ab} t^{(a} t^{b)}) = -c_2(R) \chi_R(U)
$$
we see that (\ref{QUAD}) is the correct answer up to terms of the order
of $O(t^{3/2})$ which drop in the continuum limit.
Using (\ref{QUAD}) as the starting point, it is straightforward to derive
the partition function  \cite{Mig,Rus}:
\begin{equation}
{\cal Z}(N,H,\lambda A) =
   \sum_R d_R^{2-2H}  e^{-{\lambda A c_2(R)\over 2N}}
\label{PFUN}
\end{equation}

\subsection {The  Partition Function of \gYM theories}
Pure \YM theory is in fact a special representative of a wide class of
2D gauge theories  which are invariant under area preserving diffeomorphisms.
These generalized \YM theories  are described by the following
generalized  partition function
\begin{equation}
   {\cal Z}(G,H, A,\Phi) = \int [DA^{\mu}][DB]
exp[ \int_{\Sigma} d^2x \sqrt{\det G_{ij}} ~tr(i B F - \Phi(B)) ]
\label{GENACT}
\end{equation}
where $F =  F^{ij}\epsilon_{ij}$ with $\epsilon_{ij}$
being the anti-symmetric tensor $\epsilon_{12} = -\epsilon_{21} = 1$.
$B$ is an auxiliary Lie-algebra valued
pseudo-scalar field.\footnote{In principle, we could perturb the
ordinary $YM_2$ with operators of the form ${1\over {g^{2k-2}}}tr(F^k)$,
without the need of an auxiliary field. We discuss the relation between
this perturbation and the perturbation by $tr(B^k)$ in appendix (B).}

We wish to generalize the substitution (\ref{QUAD}) for the plaquette action
(\ref{GENACT}):
\begin{equation}
Z_\triangle[U]=\int {\cal D}B e^{tr \{ia B F - t\Phi (B) \}}
 \stackrel{\rm ?}{\rightarrow} \Psi(U, t)
\label{GENYM}
\end{equation}
Here $B$ is a hermitian matrix and $\Phi$ is an invariant function
(invariant under $B\rightarrow U^{-1}BU$ for $U\in G$).
The quadratic case  $\Phi(X) = g^2tr (X^2)$ obviously correspond to the \YM
 theory.
We will take $\Phi$ to be of the form
\begin{equation}
\sum_{\{k_i\}} a_{\{k_i\}} \prod_i tr (X^i)^{k_i}
\label{GEN}
\end{equation}
(e.g. $tr (X^3)^2 + tr(X^6)$)
For $SU(N)$ ($U(N)$), $tr(X^i)$  can be expressed for $i\ge N$ ($i> N$)
in terms of $tr(X^i)$ for smaller $i$-s. Thus the summands
 in (\ref{GEN}) are not independent. This does not affect the following
discussion. Moreover, in the large $N$ limit that we will discuss
in the following section, the terms do become independent.

We define the general structure constants $d_{abc\dots k}$ to be
$$
d_{abc\dots k} \stackrel{\rm def}{=}
 g_{aa'}g_{bb'}g_{cc'}\cdots g_{kk'}
 tr (t^{a'} t^{b'} t^{c'} \cdots t^{k'})
$$

For every partition $r_1+r_2+\cdots+r_j$, we define the Casimir
\begin{eqnarray}
C_{\{r_1+r_2+\cdots+r_j\}} \stackrel{def}{=}&& \nonumber\\
{1\over{(r_1+r_2+\cdots+r_j)!}} d_{a_1^{(1)}\dots a_{r_1}^{(1)}}
d_{a_1^{(2)}\dots a_{r_2}^{(2)}} \cdots d_{a_1^{(j)}\dots a_{r_j}^{(j)}}
&& t^{(a_1^{(1)}}\cdots t^{a_{r_1}^{(1)}}
t^{a_1^{(2)}}\cdots t^{a_{r_j}^{(j)})}
\nonumber\\
&&
\label{CASIM}
\end{eqnarray}
Note that the index of $C_{\{\cdot\}}$ will always pertain to a partition.
Thus $C_{\{p\}} \neq C_{\{r_1+r_2+\cdots+r_j\}}$ even if
$p=r_1+r_2+\cdots+r_j$.
The brackets in the $t$-s mean a total symmetrization ( $(r_1+r_2+\cdots+r_j)!$
terms).

$C_\rho$ can easily be seen to commute with all the group elements and
so, by Schur's lemma, is a constant matrix in every irreducible representation.

We claim that the correct lattice generalization of (\ref{QUAD}) is
$$
\sum_R d_R \chi_R(U) e^{-t \Lambda(R)}
$$
where
\begin{equation}
\Lambda(R) = \sum_{\{k_i\}} a_{\{k_i\}} C_{\{k_1\cdot 1 + k_2 \cdot 2
+k_3 \cdot 3+\cdots \}}(R)
\label{LAMR}
\end{equation}
This results from the requirements that  $\Psi(U, t)$ must satisfy:
\begin{enumerate}
\item
$\Psi(U,0) = \delta(U-1)$.

\item $\Psi$ is a class-function.

\item
$\Psi$ satisfies the equation
$$
\{ {\partial\over {\partial t}} -\sum_{\{k_j\}} a_{\{k_j\}} \prod_l
((ia)^{-l} d_{a_1 a_2\dots a_l}{{\partial^l}\over
  {\partial F_{a_1} \cdots \partial F_{a_l}}})^{k_l} \} \Psi(U,t) +O(U-1)= 0
$$
For the $U$-s that are important in the weight for a single plaquette,
$U-1$ is of the order of magnitude of $aF$ which, in turn, is of the
order of magnitude of $O(t^{1/\nu})$ where $\nu$ is the maximal degree of
$\Phi$. Thus, the corrections to $\Psi$ are $O(a^{-(1+1/\nu)})$ and
drop out in the continuum limit.
\end{enumerate}

The partition function for the generalized $YM_2$ theory is therefore:
\begin{equation}
{\cal Z}(G,\Sigma_H,\Phi) = \sum_{R}(\dim R)^{2-2H}
e^{-{{\lambda A}\over {2N}} \Lambda(R)}\label{gPFUN}
\end{equation}
where $\Lambda(R)$ is defined in (\ref{LAMR}).
\subsection{Loop Averages in the Generalized Case}
The full solution of the \YM theory  includes in addition to the
 partition function also closed expressions for the expectation values of
products of  any
arbitrary  number of Wilson loops
$$ W(R_1,\gamma_1,...R_n\gamma_n) =<\prod_{i=1}^n Tr_{R_i}{\cal P}
e^{i\oint_{\gamma_i} A dx}>$$
where the path-ordered product around the closed curve $\gamma_i$
is taken in
the representation $R_i$. Using loop equations, derived in \cite{MakMig},
an algorithm to compute Wilson loops on the plane
was written down\cite{Kaza}. These results were derived also in
\cite{Barlic} using a different approach. Recently a prescription for
computing those averages  for non-intersecting loops on an arbitrary two
manifold  was proposed\cite{Rus}.
Let us briefly summarize the latter.
One cuts the 2D surface along the  Wilson loop contours forming several
connected ``windows". Each window contributes a sum over all irreducible
representations of the form of (\ref{PFUN}). In addition, for each pair of
neighbouring windows, a Wigner coefficient $D_{R_1R_2f}=\int dU \chi_{R_1}(U)
\chi_{R_2}(U^\dagger) \chi_{f}(U) $  is attached. Altogether,  one finds
$$ W(R_1,\gamma_1,...R_n\gamma_n) ={1\over {\cal Z}}{1\over N^n}
\sum_{R_1}....\sum_{R_n}D_{R_1...R_n}\prod_{i=1}^{N_w} d_{R_i}^{2-2G_i}
 e^{-\lambda A_i C_2(R_i)\over 2N} $$ where $N_w$  is the number of windows,
$2-2G_i$ is the Euler number associated with the window $i$
and  $D_{R_1...R_n}$ is the product of the Wigner coefficients\cite{Rus}
For the case of intersecting loops a set of differential equations \cite{Kaza}
provides a recursion relation by relating the average of a loop with $n$
intersections to those of loops with $m<n$ intersections.

 Generalizing these results to the \gYM is straightforward. The only
alteration that has to be invoked is to replace the $e^{-\lambda A C_2(R)\over
2N}$ factors that show up in those algorithms with similar factor where the
second Casimir operator is replaced by the
generalized Casimir operator (\ref{LAMR}).
For instance the expectation value of a simple Wilson loop on the plane
is given by
$$ \bra W(R,\gamma)\ket = e^{-\lambda A_\gamma \Lambda(R)\over 2N} $$
where $A_\gamma$ is the  area enclosed by $\gamma$.

It is interesting to note that for odd Casimir operators the
expectation values of real representations
(like the adjoint representation)  equal unity due to the
fact that   the corresponding Casimirs vanish.
\section{Stringy \gYM- A  Gross-Taylor Type Analysis}
Following the discovery of the
 stringy nature of pure \YM theory \cite{GT,GT2,GT3,GT4}
one may anticipate that a similar situation prevails also for the
generalized \YM theories. Indeed, in this section we prove this conjecture.
Several examples serve to  demonstrate the nature of the maps that
contribute to the \gYM and their weights.
We start by reviewing the analysis of \cite{GT3,GT4}.

\subsection{ Stringy \YM theory}
The partition function expressed as a sum over
irreducible $SU(N)$ ($U(N)$) representations (\ref{PFUN}) was expanded
in \cite{GT3,GT4} in terms of powers of ${1\over N}$.
This involved expanding the  dimension and the second Casimir
operator of the various representations.
Using the Frobenius relations between representations
of the symmetric group $S_n$ and representations of $SU(N)$ ($U(N)$),
the coefficients of this asymptotic expansion were written in terms
of characters of $S_n$. The latter were further shown to correspond to
permutations of the sheets covering the target space.
 The result of  \cite{GT4} takes the form of
(using the notation of \cite{CMR})
\begin{eqnarray}
\lefteqn{
Z( A, H, N) \sim
\sum_{n^\pm,i^\pm=0}^{\infty}
\sum_{p^\pm_1,\ldots,p^\pm_{i^\pm} \in T_2\subset S_{n^\pm}}
\sum_{s^\pm_1,t^\pm_1,\ldots,s^\pm_H,t^\pm_H\in S_{n^\pm}}
\bigl({1\over N}\bigr)^{( n^+ + n^- ) (2H-2)+(i^+ + i^- )}
}\nonumber\\
&&
{{(-1)^{(i^++ i^-)}} \over  {i^+! i^-! n^+ ! n^- !}}
( \lA)^{(i^+ + i^-)}
e^{-\hlf  (n^+ + n^-) \lA}
e^{\hlf ((n^+)^2 + ( n^- )^2 - 2 n^+ n^- ) \lA/N^2}
\nonumber\\
&&
\delta_{{\scriptscriptstyle S_{n^+} \times S_{n^-}}}
\biggl ( p^+_1\cdots p^+_{i^+} p^-_1\cdots p^-_{i^- }
\Omega_{n^+, n^-}^{2-2H}
\prod_{j=1}^H [ s^+_j, t^+_j ] \prod_{k=1}^H [ s^-_k, t^-_k ]
\biggr ) \nonumber\\
\label{FULLGT}
\end{eqnarray}
where $[s,t]= sts^{-1}t^{-1}$.
Here $\delta$ is the delta function on the group algebra
of the product of symmetric groups $S_{n^+}\times S_{n^-}$,
$T_2$ is the class of elements of $S_{n^\pm}$ consisting of
 transpositions, and $\Omega^{-1}_{n^+, n^-}$ are
certain elements of the group algebra of the symmetric
group $S_{n^+} \times S_{n^-}$.

The formula (\ref{FULLGT})
nearly factorizes, splitting into a sum over $n^+,i^+,\cdots $ and
$n^-,i^-,\cdots $.
The contributions of
the $(+)$ and $(-)$
 sums were  interpreted  as arising from two ``sectors''
of a hypothetical worldsheet theory. These sectors
correspond to orientation reversing and preserving maps, respectively.
One views the $n^+=0$ and $n^-=0$ terms as
leading order terms in a $1/N$ expansion.
At higher orders the two sectors are coupled via the
$n^+ n^-$ term in the exponential and via terms in
$\Omega_{n^+ n^-}$.

Thus, the conventional $YM_2$ theory has an interpretation in terms of sums
of covering maps of the target-space. Those maps are weighted
by the factor of $N^{2-2h}e^{-\hlf n\lambda A}$ where $h$ is the genus of the
world-sheet and $A$ is the area of the target space.
The power of $N^{2-2h}$ was obtained in \cite{GT} from the Riemann-Hurwitz
formula
\begin{equation}
 2h-2 = ( n^+ + n^- ) (2H-2)+(i^+ + i^- )
\label{Hur}
\end{equation}
where $B=i^+ + i^-$ is
the total branching number.
The number of sheets above each point in target space
(the degree of the map) is $n$ and $\lambda =g^2 N$
is the string tension. Maps that have branch points are weighted by a factor
of $\lambda A$.
The dependence on the area $A$ results from the fact that the branch point
can be at any point in the target space.

\subsection{ Toward the Stringy generalized \YM}
Note that the stringy description of the \YM theory  does not attribute
 any  special weight for maps that have
 branch points of degree higher than one, nor is there a special weight
 for two (or more) branch points that
 are at the same point in the target space. The latter maps are counted with
 weight zero\cite{GT3,GT4},  at least for a toroidal
 target space, since they constitute the boundary of map space (see also
 \cite{CMR}).

The main idea behind a stringy behavior of the \gYM  is to associate
 nonzero weights to those boundary maps, once one considers
the general $\Phi(B)$ case rather than the $B^2$ theory.
 In other words, we anticipate that  we will have to add for the $tr(B^3)$
theory, for example,
maps  that have a branch point of degree 2 and count them, as well,
with a weight proportional to $A$.
{}From the technical point of view the emergence of the \YM
description in terms  of maps followed from a large $N$ expansion of the
dimensions and the second Casimir operators of (\ref{PFUN}). Obviously a
similar expansion of the former applies also for the generalized models and
therefore  what remains to be done is to properly
treat the Casimirs appearing in the exponents of (\ref{gPFUN}).

In \cite{GT}, the expansion of
the second Casimir operator $C_2(R)$ of a representation $R$ introduced
the branch points and the string tension contributions to the partition
function.  The $C_2(R)$ was expressed   in terms of
the eigenvalue of the sum of all the ${{n(n-1)}\over 2}$ transpositions
of $n$ elements (permutations containing a single cycle of length 2),
where $n$ is the number of boxes in $R$.
This is the outcome of formula (2.3) of \cite{GT} which we write as
\begin{equation}
C_2(R) = nN + 2\hP_{\{2\}}(R)
\label{SECCAS}
\end{equation}
where $\hP_{2}(R)$ is the value of the scalar matrix representing the
sum of transpositions $\sum_{i<j\le n} (ij)$ in the representation $R$ of
$S_n$ (the matrix commutes with all permutations and thus is scalar).
In the partition function, $C_2(R)$ was multiplied by ${{\lambda A}\over{2N}}$.
The resulting term $\hlf n\lambda A$ arises from the action and is proportional
to the  string tension. The term $\lambda A\hP_{\{2\}}$ arises from the
measure and is interpreted as the contribution of branch points to the weight
of a map.

Our task is, therefore, to express the generalized Casimirs $C_\rho$ of
(\ref{CASIM}) in terms of $\hP_{\rho'}$,
the generalizations of $\hP_{\{2\}}(R)$.
This is expressed  as
\begin{equation}
C_\rho(R) = \sum_{\rho'} \alpha_\rho^{\rho'} N^{h_\rho^{\rho'}} \hP_{\rho'}(R)
\label{RHOHP}
\end{equation}
where $\alpha_\rho^{\rho'}$ are coefficients that are independent of $R$ and
the power factors $h_\rho^{\rho'}$, as will be discussed in section 3.4,
 are adjusted so that a string picture is achieved.

The  $\hP_{\rho'}$ factors are associated with
$\rho'$  which is an arbitrary  partition of certain numbers, namely
 $$
\rho': \sum_i k_i\cdot i = \overbrace{1+1+\cdots+1}^{k_1}
                          +\overbrace{2+2+\cdots+2}^{k_2}
                          +\cdots
$$
$\hP_{\rho'}(R)$ is the product of two factors.
The first is the sum of all the permutations in $S_n$
($n$ is the number of boxes of $R$) which are in the equivalence class
that is characterized by having $k_i$ cycles of length $i$ for $i\ge 2$.
Just like the case of $\xP{2}(R)$, the matrix $\hP_{\rho'}(R)$
commutes with all permutations and thus is a scalar.
The sum is taken in the representation $R$ of $S_n$. The second factor
is
$$
{{n-\sum_{i=2} i k_i}\choose {k_1}}
$$
which will be interpreted
later as the number of ways to put $k_1$ marked points on the remaining
sheets that do not participate in the branch points.

\subsection { Diagrammatic expansion of the Casimir operators}
We now derive a diagrammatic expansion for the values of the Casimir operators
for any irreducible representation  $R$ of $U(N)$ and $SU(N)$ groups. The idea,
following \cite{GT3,GT4}, is to  connect $C_\rho(R)$ with the characters
of permutations in $S_n$. The $S_n$ representation is  described
by the same Young diagram as $R$.

The reader who is not interested in the details may skip this section
and continue from section (3.4) where the final result is presented.
(We will refer to the current section only once, in section (3.7),
but this is not crucial!)

The representation space of $S_n$ that corresponds to a certain
Young diagram $Y$ is a subspace of the formal algebra
of $S_n$ (i.e. all combinations
$\sum_{\sigma\in S_n} \lambda_\sigma\sigma$ where the $\lambda_\sigma$-s
are arbitrary coefficients).
It is generated by all elements of the form
$\Pi_R\sigma$  where $\sigma\in S_n$ is arbitrary and
$$
\Pi_R = \sum_{\tau\in C_Y, \nu\in R_Y} (-)^\tau\tau\nu.
$$
$C_Y$ is the set of all permutations that do not mix the columns
of $Y$ and $R_Y$ is the set of all permutations that do not mix the
rows of $Y$.

The corresponding $SU(N)$ representation is constructed as follows.
Tensor-multiply $n$ copies of ${\bf C}^N$ which we denote by
\begin{equation}
W\equiv W_1\otimes W_2 \otimes \cdots\otimes W_n.
\label{WDEF}
\end{equation}
The action of $S_n$ on $W$ is
$$
\sigma : w_1\otimes w_2\otimes\cdots\otimes w_n
\mapsto
 w_{\sigma(1)}\otimes w_{\sigma(2)}\otimes\cdots\otimes w_{\sigma(n)}.
$$
The representation $R$ of $SU(N)$ is
given by the subspace $Im \Pi_R \subset W$ (see e.g. \cite{TextGroup}).

Next consider Casimirs $C_{\{k\}}$ with subindices  which are
a partition that is a single element.
If $t^a_{(i)}$ denotes the matrix of the generator $t^a$ in the $i^{th}$
subspace $W_i$ then
\begin{equation}
t^a_R = \sum_{i=1}^n t^a_{(i)}
\label{TAR}
\end{equation}
and
\begin{equation}
C_{\{k\}} = {1\over{k!}}\sum_{a_1,\dots,a_k}
       Tr_F \{  t^{a_1}_F t^{a_2}_F \cdots t^{a_k}_F \}
        t^{(a_1}_R t^{a_2}_R \cdots t^{a_k)}_R
   \label{CK}
\end{equation}
where we use the normalization $tr_F(t^a t^b) = \delta^{ab}$.
Substituting (\ref{TAR}) in (\ref{CK}) we get
$$
C_{\{k\}} = \sum_j\sum_{\begin{array}{c}
                   r_1,r_2,\dots,r_j \\
                   r_1+r_2+\cdots+r_j = k
                  \end{array}}
      T^{(r_1,r_2,\dots,r_j)}
$$

where $T^{(r_1,r_2,\dots,r_j)}$ is the sum over all operators
of the form
\begin{equation}
{1\over{k!}}
\sum_{a_1,\dots,a_k} Tr_F \{  t^{a_1}_F t^{a_2}_F\cdots t^{a_k}_F \}
           t^{(a_1}_{(i_1)} t^{a_2}_{(i_2)} \cdots t^{a_k)}_{(i_k)}
\label{TFT}
\end{equation}
Among the indices $i_1,\dots,i_k$ several values appear more than once.
Moreover, there should be exactly $j$ different values among the $k$ indices
in such a way that the $l^{th}$ value  (out of $j$) appears $r_l$ times.
For every permutation $\sigma\in S_k$ define
$P_\sigma$ which acts on
$$
W_{i_1}\otimes W_{i_2} \otimes \cdots \otimes W_{i_k}
$$
by
$$
(P_\sigma)^{\a_1\a_2\dots\a_k}_{\b_1\b_2\dots\b_k}
=
\dl{\b_1}{\a_{\sigma(1)}}
\dl{\b_2}{\a_{\sigma(2)}}
\cdots
\dl{\b_k}{\a_{\sigma(k)}}
$$
where $\a_r,\b_r$ are indices of $W_{i_r}$.
For every equivalence class  (given by a partition $\rho$ of $n$) we define:
$$
\hP_\rho=\sum_{\sigma\in\rho}P_\sigma
$$
Our task is to express (\ref{TFT}) in terms of the various $\hP_\rho$-s.
For this end we define
$$
\L_r = \sum_a t^a_F t^a_{(i_r)}
$$
$\L_r$ is a matrix in $F\otimes W_{i_r}$ where $F$ is a fixed vector space
(isomorphic to the fundamental representation space).
It can be checked that, in indices,
\be
(\L_r)^{\a',\b'}_{\a,\b}
  = \dl{\a}{\b'} \dl{\b}{\a'}-\eN \dl{\a}{\a'} \dl{\b}{\b'}
\label{LRa}
\ee
where $\a$ and $\a'$ are indices of $F$, $\b$ and $\b'$ are indices
of $W_{i_r}$ and $\epsilon = 0$ for $U(N)$ and $\epsilon = 1$ for $SU(N)$.
We have
\begin{eqnarray}
\lefteqn{
\sum_{a_1,\dots,a_k} Tr_F \{  t^{a_1}_F t^{a_2}_F\cdots t^{a_k}_F \}
   (t^{a_1}_{(i_1)})_{\a_1}^{\b_1}
   (t^{a_2}_{(i_2)})_{\a_2}^{\b_2}
   \cdots
   (t^{a_k}_{(i_k)})_{\a_k}^{\b_k}
}\nonumber\\
&=&\sum_{\g_1,\g_2,\dots,\g_k}
  (\L_1)^{\g_2,\b_1}_{\g_1,\a_1}
  (\L_2)^{\g_3,\b_2}_{\g_2,\a_2}
  \cdots
  (\L_{k-1})^{\g_k,\b_{k-1}}_{\g_{k-1},\a_{k-1}}
  (\L_k)^{\g_1,\b_k}_{\g_k,\a_k}
\end{eqnarray}

Using (\ref{LRa}) we can describe this, graphically, as a sum of
diagrams.  This is demonstrated in  Fig.1 for a particular term of
$C_{\{6\}}$.

\parbox[c]{140mm}{
\begin{picture}(140,80)

\thicklines

\put(10,40){\circle{2}}
\put(20,50){\circle{2}}
\put(20,30){\circle{2}}
\put(11,41){\line(1,1){8}}
\put(29,39){\line(-1,-1){8}}
\put(16,10){$(\Lambda_1)$}
\put(15,16){$\dl{\g_1}{\b_1} \dl{\a_1}{\g_2}$}
\put(18,53){$\b_1$}
\put(18,24){$\a_1$}
\put(8,45){$\g_1$}

\put(30,40){\circle{2}}
\put(40,50){\circle{2}}
\put(40,30){\circle{2}}
\put(31,41){\line(1,1){8}}
\put(49,39){\line(-1,-1){8}}
\put(36,10){$(\Lambda_2)$}
\put(35,16){$\dl{\g_2}{\b_2} \dl{\a_2}{\g_3}$}
\put(38,53){$\b_2$}
\put(38,24){$\a_2$}
\put(28,45){$\g_2$}

\put(50,40){\circle{2}}
\put(60,50){\circle{2}}
\put(60,30){\circle{2}}
\put(51,41){\line(1,1){8}}
\put(69,39){\line(-1,-1){8}}
\put(56,10){$(\Lambda_3)$}
\put(55,16){$\dl{\g_3}{\b_3} \dl{\a_3}{\g_4}$}
\put(58,53){$\b_3$}
\put(58,24){$\a_3$}
\put(48,45){$\g_3$}

\put(73,65){{\huge $({{-\epsilon}\over N})$}}
\put(70,40){\circle{2}}
\put(80,50){\circle{2}}
\put(80,30){\circle{2}}
\put(80,31){\line(0,1){7}}
\put(80,49){\line(0,-1){7}}
\put(71,40){\line(1,0){18}}
\put(76,10){$(\Lambda_4)$}
\put(75,16){$\dl{\g_4}{\g_5} \dl{\a_4}{\b_4}$}
\put(78,53){$\b_4$}
\put(78,24){$\a_4$}
\put(68,45){$\g_4$}

\put(93,65){{\huge $({{-\epsilon}\over N})$}}
\put(90,40){\circle{2}}
\put(100,50){\circle{2}}
\put(100,30){\circle{2}}
\put(100,31){\line(0,1){7}}
\put(100,49){\line(0,-1){7}}
\put(91,40){\line(1,0){18}}
\put(96,10){$(\Lambda_5)$}
\put(95,16){$\dl{\g_5}{\g_6} \dl{\a_5}{\b_5}$}
\put(98,53){$\b_5$}
\put(98,24){$\a_5$}
\put(88,45){$\g_5$}

\put(110,40){\circle{2}}
\put(120,50){\circle{2}}
\put(120,30){\circle{2}}
\put(111,41){\line(1,1){8}}
\put(129,39){\line(-1,-1){8}}
\put(116,10){$(\Lambda_6)$}
\put(115,16){$\dl{\g_6}{\b_6} \dl{\a_6}{\g_1}$}
\put(118,53){$\b_6$}
\put(118,24){$\a_6$}
\put(108,45){$\g_6$}
\put(128,45){$\g_1$}
\put(130,40){\circle{2}}

\end{picture}
\\
{\it Fig. 1: Diagrammatic description of the multiplication
     of the $\Lambda_r$ matrices.}\\ \\ }
\\
{}From this picture we see that
\begin{eqnarray}
\lefteqn{
\sum_{\g_1,\g_2,\dots,\g_k}
  (\L_1)^{\g_2,\b_1}_{\g_1,\a_1}
  (\L_2)^{\g_3,\b_2}_{\g_2,\a_2}
  \cdots
  (\L_{k-1})^{\g_k,\b_{k-1}}_{\g_{k-1},\a_{k-1}}
  (\L_k)^{\g_1,\b_k}_{\g_k,\a_k}
} \nonumber\\
&=&
\sum_{p=0}^k (-\eN)^{k-p} \,\,\,\{
 \sum_{1 \le r_1 < r_2 < \cdots < r_p \le k} \!\!\!
 (\dl{\a_{r_1}}{\b_{r_2}}
  \dl{\a_{r_2}}{\b_{r_3}}
  \cdots
  \dl{\a_{r_{k-1}}}{\b_{r_k}}
  \dl{\a_{r_k}}{\b_{r_1}})
 \prod_{s\not\in\{r_1,\dots,r_p\}}\!\!\!\!\!\!\!\! \dl{\a_s}{\b_s}
 \,\}
\label{LRLR}
\end{eqnarray}

We will draw the permutation that corresponds to Fig.1 as:
\vskip 0.5cm
\parbox[c]{100mm}{
\begin{picture}(100,50)
\put(6,19){$3$}
\put(10,20){\circle{2}}
\put(19,32){$2$}
\put(20,30){\circle{2}}
\put(32,19){$1$}
\put(30,20){\circle{2}}
\put(19,5){$6$}
\put(20,10){\circle{2}}
\put(19,21){\oval(18,18)[tl]}
\put(19,19){\oval(18,18)[bl]}
\put(21,19){\oval(18,18)[br]}
\put(21,21){\oval(18,18)[tr]}
\put(30,18){\vector(0,1){1}}
\put(10,22){\vector(0,-1){1}}
\put(22,30){\vector(-1,0){1}}
\put(18,10){\vector(1,0){1}}

\put(45,25){$4$}
\put(46,23){\circle{2}}
\put(45,20){\oval(6,6)[l]}
\put(47,20){\oval(6,6)[r]}
\put(45,17){\line(1,0){2}}
\put(48,23){\vector(-1,0){1}}

\put(65,25){$5$}
\put(66,23){\circle{2}}
\put(65,20){\oval(6,6)[l]}
\put(67,20){\oval(6,6)[r]}
\put(65,17){\line(1,0){2}}
\put(68,23){\vector(-1,0){1}}
\end{picture}
\\
{\it Fig. 2: The permutation of Fig.1.  }\\ \\ }
\\
{}From Figs.1 and 2 we obtain immediately (see (\ref{TFT})):
\begin{equation}
T^{(\overbrace{1,1,\dots,1}^k)} =  N(-\eN)^k\hP_{\{k\cdot 1\}}
+\sum_{p=1}^k {{k!}\over {p!}}
(-\eN)^{k-p} \hP_{\{1\cdot p + (k-p)\cdot 1\}}
\end{equation}
where $\{1\cdot p + (k-p)\cdot 1\}$ is the partition
$p + \overbrace{1+1+\cdots+1}^{k-p}$ and $T^{(\overbrace{1,1,\dots,1}^k)}$
was defined in eqn. (\ref{TFT})

We can obtain the other $T^{(r_1,\dots,r_j)}$ from (\ref{LRLR})
by contracting some of the $\a_{i_l}$-s with the $\b_{i_m}$-s.
For example, for $j=1$, $T^{(k)}$ is
\begin{eqnarray}
\sum_{i=1}^n
\lefteqn{
\sum_{i=1}^n
\sum_{a_1,\dots,a_k} Tr_F \{  t^{a_1}_F t^{a_2}_F\cdots t^{a_k}_F \}
   (t^{(a_1}_{(i)})_{\a_1}^{\a_2}
   (t^{a_2}_{(i)})_{\a_2}^{\a_3}
   \cdots
   (t^{a_k)}_{(i)})_{\a_{k}}^{\a_{k+1}}
} \nonumber\\
&=&
{1\over{(k-1)!}}
\sum_{i=1}^n
\sum_{a_1,\dots,a_k} Tr_F \{  t^{a_1}_F t^{a_2}_F\cdots t^{a_k}_F \}
\!\!\!\!
\sum_{\sigma\in [\{k\}]\subset S_k}
   (t^{(a_1}_{(i)})_{\a_{\sigma(1)}}^{\a_2}
   (t^{a_2}_{(i)})_{\a_{\sigma(2)}}^{\a_3}
   \cdots
   (t^{a_k)}_{(i)})_{\a_{\sigma(k)}}^{\a_{k+1}}
\nonumber\\
&&
\end{eqnarray}
The symbol $[\{k\}]$ is the equivalence class of all the $(k-1)!$
permutations in $S_k$ that have just one cycle (of length $k$).
Graphically, we describe contractions of the
$\a_{i_l}$-s with the $\b_{i_m}$-s by directed dashed lines.
The diagrams that are obtained are of the  form drawn in the following
figure:

\vskip 0.5cm
\parbox[c]{100mm}{

\begin{picture}(100,60)

\thicklines

\put(40,18){$\cdot$}
\put(38,16){$\cdot$}
\put(35,14){$\cdot$}

\put(43,37){\circle{2}}
\put(45,37){$2$}
\put(45,31){\vector(-1,4){1}}
\put(45,30){\circle{2}}
\put(47,29){$1$}
\put(44,24){\vector(1,4){1}}
\put(43,23){\circle{2}}
\put(45,21){$k$}

\put(24,47){\circle{2}}
\put(22,49){$x$}
\put(23,47){\vector(-3,-1){4}}
\put(17,45){\circle{2}}
\put(14,46){$y$}
\put(16,44){\vector(-2,-3){3}}
\put(12,38){\circle{2}}
\put(9,39){$z$}

\put(31,45){$\cdot$}
\put(35,43){$\cdot$}
\put(38,41){$\cdot$}

\put(30,13){\circle{2}}
\put(28,8){$w'$}
\put(24,13){\vector(1,0){5}}
\put(23,13){\circle{2}}
\put(21,8){$z'$}
\put(18,14){\vector(3,-1){4}}
\put(17,15){\circle{2}}
\put(12,11){$y'$}
\put(12,19){\vector(1,-1){4}}
\put(12,20){\circle{2}}
\put(6,18){$x'$}

\put(10,32){$\cdot$}
\put(9,27){$\cdot$}
\put(10,22){$\cdot$}

\put(17,16){\line(0,1){1}}
\put(17,18){\line(0,1){1}}
\put(17,20){\line(0,1){1}}
\put(17,22){\line(0,1){1}}
\put(17,24){\line(0,1){1}}
\put(17,26){\line(0,1){1}}
\put(17,28){\line(0,1){1}}
\put(17,30){\vector(0,1){1}}
\put(17,32){\line(0,1){1}}
\put(17,34){\line(0,1){1}}
\put(17,36){\line(0,1){1}}
\put(17,38){\line(0,1){1}}
\put(17,40){\line(0,1){1}}
\put(17,42){\line(0,1){1}}

\put(13,38){\line(1,0){1}}
\put(15,38){\line(1,0){1}}
\put(18,38){\line(1,0){1}}
\put(20,38){\line(1,0){1}}
\put(22,38){\line(1,0){1}}
\put(24,38){\line(1,0){1}}
\put(26,38){\line(1,0){1}}
\put(28,38){\vector(1,0){1}}
\put(30,38){\line(1,0){1}}
\put(32,38){\line(1,0){1}}
\put(34,38){\line(1,0){1}}
\put(36,38){\line(1,0){1}}
\put(38,38){\line(1,0){1}}
\put(40,38){\line(1,0){1}}

\put(12,21){\line(0,1){1}}
\put(12,23){\line(0,1){1}}
\put(12,25){\line(0,1){1}}
\put(12,27){\line(0,1){1}}
\put(12,29){\vector(0,1){1}}
\put(12,31){\line(0,1){1}}
\put(12,33){\line(0,1){1}}
\put(12,35){\line(0,1){1}}

\end{picture}
\\
{\it Fig. 3: An example of a ``dashed'' diagram.} \\ \\ }
\\
There are no two incoming dashed arrows or two outcoming dashed arrows
from the same vertex `$\circ$'.

The dashed lines can be eliminated by two rules  described in  Fig.4a
and Fig.4b.  This can   be checked with the use of  Fig.1.

\vskip 0.5cm
\parbox[c]{140mm}{
\begin{picture}(140,60)

\thicklines

\put(40,18){$\cdot$}
\put(38,16){$\cdot$}
\put(35,14){$\cdot$}

\put(43,37){\circle{2}}
\put(45,37){$2$}
\put(45,31){\vector(-1,4){1}}
\put(45,30){\circle{2}}
\put(47,29){$1$}
\put(44,24){\vector(1,4){1}}
\put(43,23){\circle{2}}
\put(45,21){$k$}

\put(24,47){\circle{2}}
\put(22,49){$x$}
\put(23,47){\vector(-3,-1){4}}
\put(17,45){\circle{2}}
\put(14,46){$y$}
\put(16,44){\vector(-2,-3){3}}
\put(12,38){\circle{2}}
\put(9,39){$z$}

\put(31,45){$\cdot$}
\put(35,43){$\cdot$}
\put(38,41){$\cdot$}

\put(30,13){\circle{2}}
\put(28,8){$w'$}
\put(24,13){\vector(1,0){5}}
\put(23,13){\circle{2}}
\put(21,8){$z'$}
\put(18,14){\vector(3,-1){4}}
\put(17,15){\circle{2}}
\put(12,11){$y'$}
\put(12,19){\vector(1,-1){4}}
\put(12,20){\circle{2}}
\put(6,18){$x'$}

\put(10,32){$\cdot$}
\put(10,27){$\cdot$}
\put(10,22){$\cdot$}

\put(17,16){\line(0,1){1}}
\put(17,18){\line(0,1){1}}
\put(17,20){\line(0,1){1}}
\put(17,22){\line(0,1){1}}
\put(17,24){\line(0,1){1}}
\put(17,26){\line(0,1){1}}
\put(17,28){\line(0,1){1}}
\put(17,30){\vector(0,1){1}}
\put(17,32){\line(0,1){1}}
\put(17,34){\line(0,1){1}}
\put(17,36){\line(0,1){1}}
\put(17,38){\line(0,1){1}}
\put(17,40){\line(0,1){1}}
\put(17,42){\line(0,1){1}}

\put(55,31){\line(1,0){8}}
\put(55,29){\line(1,0){8}}


\put(77,37){\circle{2}}
\put(82,34){\vector(-3,2){4}}
\put(83,33){\circle{2}}
\put(84,35){$z$}
\put(83,28){\vector(0,1){4}}
\put(83,27){\circle{2}}
\put(84,24){$y=y'$}
\put(78,23){\vector(3,2){4}}
\put(77,23){\circle{2}}
\put(76,18){$x'$}

\put(71,23){$\cdot$}
\put(69,25){$\cdot$}
\put(68,28){$\cdot$}
\put(69,31){$\cdot$}
\put(71,34){$\cdot$}

\put(132,37){\circle{2}}
\put(134,37){$2$}
\put(135,31){\vector(-1,3){2}}
\put(135,30){\circle{2}}
\put(137,29){$1$}
\put(132,23){\vector(1,2){3}}
\put(131,22){\circle{2}}
\put(133,20){$k$}

\put(112,40){$\cdot$}
\put(115,42){$\cdot$}
\put(118,43){$\cdot$}
\put(121,43){$\cdot$}
\put(124,42){$\cdot$}
\put(127,40){$\cdot$}

\put(108,37){\circle{2}}
\put(105,38){$x$}
\put(107,36){\vector(-1,-2){2}}
\put(105,30){\circle{2}}
\put(100,29){$z'$}
\put(105,29){\vector(1,-2){3}}
\put(109,22){\circle{2}}
\put(105,17){$w'$}

\put(112,18){$\cdot$}
\put(115,16){$\cdot$}
\put(118,15){$\cdot$}
\put(121,15){$\cdot$}
\put(124,16){$\cdot$}
\put(127,18){$\cdot$}

\end{picture}
\\
{\it Fig. 4a: Rules for eliminating dashed lines.  }\\ \\ }
\\
\vskip 0.5cm
\parbox[c]{140mm}{
\begin{picture}(140,60)

\thicklines

\put(130,18){$\cdot$}
\put(128,16){$\cdot$}
\put(125,14){$\cdot$}
\put(122,13){$\cdot$}
\put(118,12){$\cdot$}

\put(133,37){\circle{2}}
\put(135,37){$2$}
\put(135,31){\vector(-1,4){1}}
\put(135,30){\circle{2}}
\put(137,29){$1$}
\put(134,24){\vector(1,4){1}}
\put(133,23){\circle{2}}
\put(135,21){$k$}

\put(114,47){\circle{2}}
\put(112,49){$x'$}
\put(113,47){\vector(-3,-1){4}}
\put(107,45){\circle{2}}
\put(104,46){$z$}

\put(120,46){$\cdot$}
\put(124,45){$\cdot$}
\put(127,43){$\cdot$}
\put(130,40){$\cdot$}

\put(113,13){\circle{2}}
\put(111,8){$w'$}
\put(108,14){\vector(3,-1){4}}
\put(107,15){\circle{2}}
\put(96,11){$y=z'$}
\put(102,19){\vector(1,-1){4}}
\put(102,20){\circle{2}}
\put(97,17){$x$}

\put(102,40){$\cdot$}
\put(100,37){$\cdot$}
\put(99,32){$\cdot$}
\put(99,27){$\cdot$}
\put(100,22){$\cdot$}

\put(83,31){\line(1,0){8}}
\put(83,29){\line(1,0){8}}


\put(21,36){\circle{2}}
\put(23,38){$z$}
\put(24,31){\vector(-1,2){2}}
\put(24,30){\circle{2}}
\put(26,34){$y$}
\put(22,25){\vector(1,2){2}}
\put(21,24){\circle{2}}
\put(20,19){$x$}

\put(17,22){$\cdot$}
\put(14,22){$\cdot$}
\put(11,23){$\cdot$}
\put(9,25){$\cdot$}
\put(8,28){$\cdot$}
\put(9,31){$\cdot$}
\put(11,34){$\cdot$}
\put(14,36){$\cdot$}
\put(17,36){$\cdot$}

\put(25,30){\line(1,0){1}}
\put(27,30){\line(1,0){1}}
\put(29,30){\line(1,0){1}}
\put(31,30){\line(1,0){1}}
\put(33,30){\line(1,0){1}}
\put(35,30){\vector(1,0){1}}
\put(37,30){\line(1,0){1}}
\put(39,30){\line(1,0){1}}
\put(41,30){\line(1,0){1}}
\put(43,30){\line(1,0){1}}

\put(72,37){\circle{2}}
\put(74,37){$2$}
\put(75,31){\vector(-1,3){2}}
\put(75,30){\circle{2}}
\put(77,29){$1$}
\put(72,23){\vector(1,2){3}}
\put(71,22){\circle{2}}
\put(73,20){$k$}

\put(52,40){$\cdot$}
\put(55,42){$\cdot$}
\put(58,43){$\cdot$}
\put(61,43){$\cdot$}
\put(64,42){$\cdot$}
\put(67,40){$\cdot$}

\put(48,37){\circle{2}}
\put(45,38){$x'$}
\put(47,36){\vector(-1,-2){2}}
\put(45,30){\circle{2}}
\put(40,33){$z'$}
\put(45,29){\vector(1,-2){3}}
\put(49,22){\circle{2}}
\put(45,17){$w'$}

\put(52,18){$\cdot$}
\put(55,16){$\cdot$}
\put(58,15){$\cdot$}
\put(61,15){$\cdot$}
\put(64,16){$\cdot$}
\put(67,18){$\cdot$}

\end{picture}
\\
{\it Fig. 4b: Rules for eliminating dashed lines.  }\\ \\ }
\\
The next step is to consider trajectories of dashed lines.
Suppose that such a trajectory passes through $r$ vertices.
Then there are $r!$ different ways to draw directed dashed lines that
pass through those $r$ vertices in a single trajectory.
We will first sum over all those $r!$ dashed lines trajectories, as
in Fig.5.
\vskip 0.5cm
\parbox[c]{150mm}{
\begin{picture}(150,70)

\thicklines

\put(120,35){\circle{2}}
\put(121,43){\circle{2}}
\put(130,45){\circle{2}}
\put(139,43){\circle{2}}
\put(140,35){\circle{2}}

\put(141,36){\vector(-1,3){2}}
\put(138,43){\vector(-3,1){6}}
\put(129,45){\vector(-3,-1){6}}
\put(121,42){\vector(-1,-3){2}}

\put(130,21){$\cdot$}
\put(125,23){$\cdot$}
\put(135,23){$\cdot$}
\put(122,27){$\cdot$}
\put(138,27){$\cdot$}

\thinlines
\put(118,36){\line(1,1){11}}
\put(142,36){\line(-1,1){11}}
\put(118,33){\line(1,0){24}}

\put(129,47){\line(1,0){2}}
\put(118,36){\line(0,-1){3}}
\put(142,36){\line(0,-1){3}}

\thicklines
\put(97,35){\line(1,0){8}}
\put(97,33){\line(1,0){8}}

\thicklines

\put(5,20){\circle{2}}
\put(6,28){\circle{2}}
\put(15,30){\circle{2}}
\put(24,28){\circle{2}}
\put(25,20){\circle{2}}

\put(26,21){\vector(-1,3){2}}
\put(23,28){\vector(-3,1){6}}
\put(14,30){\vector(-3,-1){6}}
\put(6,27){\vector(-1,-3){2}}

\put(14,7){$\cdot$}
\put(9,9){$\cdot$}
\put(19,9){$\cdot$}
\put(6,13){$\cdot$}
\put(22,13){$\cdot$}

\thinlines
\put(5,20){\line(1,1){4}}
\put(11,26){\vector(1,1){4}}
\put(15,30){\line(1,-1){4}}
\put(21,24){\vector(1,-1){4}}

\thicklines

\put(35,20){\circle{2}}
\put(36,28){\circle{2}}
\put(45,30){\circle{2}}
\put(54,28){\circle{2}}
\put(55,20){\circle{2}}

\put(56,21){\vector(-1,3){2}}
\put(53,28){\vector(-3,1){6}}
\put(44,30){\vector(-3,-1){6}}
\put(36,27){\vector(-1,-3){2}}

\put(44,7){$\cdot$}
\put(39,9){$\cdot$}
\put(49,9){$\cdot$}
\put(36,13){$\cdot$}
\put(52,13){$\cdot$}

\thinlines
\put(35,20){\line(1,1){4}}
\put(41,26){\vector(1,1){4}}
\put(54,20){\line(-1,0){4}}
\put(48,20){\line(-1,0){4}}
\put(42,20){\vector(-1,0){4}}

\thicklines

\put(65,20){\circle{2}}
\put(66,28){\circle{2}}
\put(75,30){\circle{2}}
\put(84,28){\circle{2}}
\put(85,20){\circle{2}}

\put(86,21){\vector(-1,3){2}}
\put(83,28){\vector(-3,1){6}}
\put(74,30){\vector(-3,-1){6}}
\put(66,27){\vector(-1,-3){2}}

\put(74,7){$\cdot$}
\put(69,9){$\cdot$}
\put(79,9){$\cdot$}
\put(66,13){$\cdot$}
\put(82,13){$\cdot$}

\thinlines
\put(75,30){\line(1,-1){4}}
\put(81,24){\vector(1,-1){4}}
\put(84,20){\line(-1,0){4}}
\put(78,20){\line(-1,0){4}}
\put(72,20){\vector(-1,0){4}}

\thicklines

\put(5,50){\circle{2}}
\put(6,58){\circle{2}}
\put(15,60){\circle{2}}
\put(24,58){\circle{2}}
\put(25,50){\circle{2}}

\put(26,51){\vector(-1,3){2}}
\put(23,58){\vector(-3,1){6}}
\put(14,60){\vector(-3,-1){6}}
\put(6,57){\vector(-1,-3){2}}

\put(14,37){$\cdot$}
\put(9,39){$\cdot$}
\put(19,39){$\cdot$}
\put(6,43){$\cdot$}
\put(22,43){$\cdot$}

\thinlines
\put(24,51){\line(-1,1){4}}
\put(19,56){\vector(-1,1){4}}
\put(15,60){\line(-1,-1){4}}
\put(9,54){\vector(-1,-1){4}}

\thicklines

\put(35,50){\circle{2}}
\put(36,58){\circle{2}}
\put(45,60){\circle{2}}
\put(54,58){\circle{2}}
\put(55,50){\circle{2}}

\put(56,51){\vector(-1,3){2}}
\put(53,58){\vector(-3,1){6}}
\put(44,60){\vector(-3,-1){6}}
\put(36,57){\vector(-1,-3){2}}

\put(44,37){$\cdot$}
\put(39,39){$\cdot$}
\put(49,39){$\cdot$}
\put(36,43){$\cdot$}
\put(52,43){$\cdot$}

\thinlines
\put(36,50){\line(1,0){4}}
\put(42,50){\line(1,0){4}}
\put(48,50){\vector(1,0){4}}
\put(45,60){\line(-1,-1){4}}
\put(39,54){\vector(-1,-1){4}}

\thicklines

\put(65,50){\circle{2}}
\put(66,58){\circle{2}}
\put(75,60){\circle{2}}
\put(84,58){\circle{2}}
\put(85,50){\circle{2}}

\put(86,51){\vector(-1,3){2}}
\put(83,58){\vector(-3,1){6}}
\put(74,60){\vector(-3,-1){6}}
\put(66,57){\vector(-1,-3){2}}

\put(74,37){$\cdot$}
\put(69,39){$\cdot$}
\put(79,39){$\cdot$}
\put(66,43){$\cdot$}
\put(82,43){$\cdot$}

\thinlines
\put(84,51){\line(-1,1){4}}
\put(79,56){\vector(-1,1){4}}
\put(66,50){\line(1,0){4}}
\put(72,50){\line(1,0){4}}
\put(78,50){\vector(1,0){4}}

\thicklines
\put(60,18){\line(0,1){4}}
\put(58,20){\line(1,0){4}}

\put(30,18){\line(0,1){4}}
\put(28,20){\line(1,0){4}}

\put(60,48){\line(0,1){4}}
\put(58,50){\line(1,0){4}}

\put(30,48){\line(0,1){4}}
\put(28,50){\line(1,0){4}}

\put(3,33){\line(0,1){4}}
\put(1,35){\line(1,0){4}}

\end{picture}
\\
{\it Fig. 5: Summing over dashed trajectories.} \\ \\ } \\

Using the rules of $Fig.4a-b$ and induction on $r$, we
easily find the rule of Fig.6.

\vskip 0.5cm
\parbox[c]{140mm}{
\begin{picture}(140,50)

\put(25,20){\oval(26,10)}

\put(17,20){\circle{2}}
\put(16,21){\vector(-1,1){8}}
\put(8,11){\vector(1,1){8}}
\put(7,10){\circle{2}}
\put(3,6){$i_1$}
\put(7,30){\circle{2}}
\put(3,33){$j_1$}

\put(21,20){\circle{2}}
\put(21,21){\vector(-1,2){5}}
\put(16,9){\vector(1,2){5}}
\put(16,8){\circle{2}}
\put(15,3){$i_2$}
\put(16,32){\circle{2}}
\put(15,35){$j_2$}

\put(25,19){$\cdots$}

\put(33,20){\circle{2}}
\put(34,21){\vector(1,1){8}}
\put(42,11){\vector(-1,1){8}}
\put(43,10){\circle{2}}
\put(44,6){$i_r$}
\put(43,30){\circle{2}}
\put(44,33){$j_r$}

\put(48,21){\line(1,0){8}}
\put(48,19){\line(1,0){8}}

\put(60,18){\Huge ${1\over{r!}}\,\,\,\,
             \Sigma
              \,\,\,\,\,\,\,
             \Sigma
              \,\,
\left\{
\begin{array}{c} \,\,\,\,\,\,\,\,\,\,\,\,\,\,\,\,\,\,\,\,\,\,\,\,\,\,\,\, \\
                 \,\,\,\,\,\,\,\,\,\,\,\,\,\,\,\,\,\,\,\,\,\,\,\,\,\,\,\,
\end{array}\right\}$}
\put(67,13){$\sigma\in [\{r\}]$}
\put(86,13){$s\!\!=\!\!1$}
\put(89,25){$r$}

\put(101,32){\circle{2}}
\put(101,9){\vector(0,1){22}}
\put(101,8){\circle{2}}
\put(100,3){$i_1$}
\put(96,36){$j_{\sigma(1)}$}

\put(107,32){\circle{2}}
\put(107,9){\vector(0,1){22}}
\put(107,8){\circle{2}}
\put(106,3){$i_2$}
\put(106,36){$j_{\sigma(2)}$}

\put(111,19){$\cdots$}

\put(121,20){\circle{2}}
\put(121,21){\vector(0,1){10}}
\put(121,32){\circle{2}}
\put(121,9){\vector(0,1){10}}
\put(121,8){\circle{2}}
\put(120,3){$i_s$}
\put(120,36){$j_{\sigma(s)}$}

\put(125,19){$\cdots$}

\put(133,32){\circle{2}}
\put(133,9){\vector(0,1){22}}
\put(133,8){\circle{2}}
\put(132,3){$i_r$}
\put(132,36){$j_{\sigma(r)}$}

\end{picture}
\\
{\it Fig. 6: Graphical description of the contraction of $r$ vertices.
     The notation $\sigma\in [\{r\}]$ means that $\sigma$ is a permutation
     with just one cycle of length $r$.
     Note that the number of terms on the rhs is exactly $r!$, since
     there are $(r-1)!$ permutations in the equivalence class $[\{r\}]$.
     }\\ \\ } \\

At this stage the diagrammatic expansion of the Casimirs includes
 a sum of diagrams, each composed of loops of various
sizes. Each loop of size zero should be replaced by an extra factor
of $N$ to its corresponding diagram.
We now look at a specific diagram. Suppose it has $\mu_t$ loops of size $t$,
for $t=1,2,\dots$. Let $m=\sum_{t=2}^\infty t\mu_t$.
The number of ways to write indices from $1,2,\dots,n$ where
$n$ is the number of boxes in the representation, or as we discuss later, the
number of sheets in the string picture, is ${{n-m}\choose {n-m-\mu_1}}$.
If $m+\mu_1 > n$ then, by definition, this factor is zero.
Since $m$ is independent of $n$, and in fact depends only on $\Phi(B)$,
the stringy interpretation of this factor is of $\mu_1$
microscopic holes that we cut in the sheet, above the singular point.

Thus, when we discuss
the weight that is to be associated with a branch point that makes a
permutation of the sheets which has $k_i$ cycles of length $i\ge 2$,
(for example, for a simple branch point $k_2=2$ and all the other
$k_i$-s are zero) it is natural to include $i=1$ as well.
We interpret such points as places where there are holes in $k_1$ sheets
(possibly in addition to other branch points which mix {\em other} sheets,
if $k_i\ne 0$ for other $i$-s).

The weight that is to be associated to such a generalized branch point is
$\lambda A$ times the sum of the coefficients of all the diagrams that
have $k_i$ loops of length $i$ (after the manipulations of Fig.6), times
a symmetry factor of $\prod_{i=1}^\infty (i^{k_i}k_i!)$.
This factor arises from the $k_i!$ possible ways to mix the loops of size
$i$, each corresponding to a different indexing but the same permutation.
Also, there are $i$ ways to rotate each loop, which again corresponds to
a different indexing (from $1,\dots,n$) but the same permutation.

Note that the factor of ${1\over {r!}}$ in Fig.5 arises from the fact that
there is a symmetry factor of ${{k!}\over{r_1! r_2!\cdots r_j!}}$ in
(\ref{TFT}). Indeed, if we permute the subindices $i_1,\dots,i_k$
and the indices $a_1,\dots,a_k$ of (\ref{TFT}) in such a way that
we do not change the relative order of the $a_q$-s
among $t^{a_q}_{(i_q)}$-s that have the same $i_q$-s
(because they multiply each other in the same vector space $W_{i_q}$),
we do not change the graph of the type of Fig.3 that corresponds to it.
The factor of $k!$ cancels with the factor of ${1\over{k!}}$ in
(\ref{TFT}).

The generalization to the non-chiral sector is done by replacing
(\ref{WDEF}) by
\begin{equation}
W\equiv W_1\otimes W_2 \otimes \cdots\otimes W_{n^+}
  \otimes \bW_1\otimes \bW_2 \otimes \cdots\otimes \bW_{n^-}
\end{equation}
where a generator $t^a$  acts on $\bW_i$ as the transposed matrix of $t^a$
in the fundamental representation.
The representation $R\bar{S}$ (following \cite{GT}) of $SU(N)$ (or $U(N)$)
is obtained by restricting to the kernel of all the operators that contract
an index of one of the $\bW_i$-s with an index of one of the $W_j$-s,
and also applying the two projection operators $\Pi_R$ and $\Pi_S$,
where $R$ is the corresponding Young diagram for $S_{n^+}$ and $S$ is the Young
diagram for $S_{n^-}$.

Putting everything together we have the following rules to obtain the
cycle structure:

\subsection{Cycle structure formula for $U(N)$ and $SU(N)$}
The coefficients of the  various $\hP_{\rho'}(R)$ which soon will be shown
to correspond to
the weight for a map with a point that has a multiple branch point
(and possibly multiple marked points) is obtained diagrammatically
as follows:
\begin{enumerate}
\item
For every term in $\Phi(B)$ of the form $\prod_k \{tr(B^k)\}^{\nu_k}$
draw a graph of $\sum\nu_k$ loops so that there are $\nu_k$ loops of
size $k$ (and each loop is oriented).

\item
Divide the various loops into two groups, in all possible ways.
The first group will be called the {\em chiral} group and the second
group will be called the {\em anti-chiral} group.
Each diagram will have a factor of
$$
(-)^{\mbox{(number of anti-chiral vertices)}}.
$$

\item
Group the $k\nu_k$ vertices of all the loops into
groups in all possible ways with the restriction
that vertices from chiral and anti-chiral
loops do not belong to the same group. Each group corresponds to one ellipse
as in Fig.6 (where the group is of size $r$).

\item
Reduce each group to one vertex according to the rule of Fig.6.

\item
The result is a sum of diagrams, each composed of loops of various
sizes. Each loop of size zero
contributes a factor
$N$ to its corresponding diagram.
\end{enumerate}

For $SU(N)$  we have to include the $-{1\over N}$ factors from Fig.7.
The rules are exactly the same as for the $U(N)$ case above, except that
after we have drawn the diagrams that correspond to the terms in
$\Phi(B)$ (step (1) above) we have to add all possible diagrams that are
obtained from the previous ones by changing vertices to $1$-loops
in all possible ways. This is represented diagrammatically as:
\vskip 0.2cm
\parbox[c]{100mm}{
\begin{picture}(100,50)

\put(10,20){\circle{2}}
\put(10,21){\vector(0,1){10}}
\put(10,9){\vector(0,1){10}}
\put(9,5){$i$}
\put(9,34){$j$}

\put(20,18){$\line(1,0){8}$}
\put(20,22){$\line(1,0){8}$}
\put(30,20){$\line(-1,1){4}$}
\put(30,20){$\line(-1,-1){4}$}

\put(40,20){\circle{2}}
\put(40,21){\vector(0,1){10}}
\put(40,9){\vector(0,1){10}}
\put(39,5){$i$}
\put(39,34){$j$}

\put(50,18){\Huge $-{1\over N}
\left\{
\begin{array}{c} \,\,\,\,\,\,\,\,\,\,\,\,\, \\
                 \,\,\,\,\,\,\,\,\,\,\,\,\,
\end{array}\right\}$}

\put(71,9){\vector(0,1){22}}
\put(70,5){$i$}
\put(70,34){$j$}

\put(79,20){\circle{2}}
\put(78,17){\oval(6,6)[l]}
\put(80,17){\oval(6,6)[r]}
\put(78,14){\line(1,0){2}}
\put(81,20){\vector(-1,0){1}}

\end{picture}
\\
{\it Fig. 7: The modification from $U(N)$ to $SU(N)$} \\ \\ }

\subsection{The Partition Function as  a Sum of Maps in the Large $N$ Limit}
Equipped with the expansion of  $C_\rho$ in terms of the $\hP_{\rho'}$ factors
we proceed now to deduce the  full description of the partition function
(\ref{gPFUN}) as a sum of branched maps.
Following the same steps as in \cite{GT}, in the large $N$ limit,
the coefficients of the $\hP_{\rho'}$-s will
turn out to be the weights that are associated with a  singular point
in target space that has $k_1$ marked points, $k_2$ simple branch points,
$k_3$ branch points of degree 2, and so on
($k_i$ branch points of degree $i-1$). The numbers $k_i$ are related to the
partition $\rho'$ as above.
The only place that is different from \cite{GT} is that the
restriction $p^\pm_1,\ldots,p^\pm_{i^\pm} \in T_2\subset S_{n^\pm}$
in (\ref{FULLGT}) will be replaced by permutations that are
in the equivalence class (in $S_{n^\pm}$)
that corresponds to the partition $\rho'$ (instead of $T_2$ which corresponds
to the simple partition $\{2\}$).

Using the cycle structure formula  of the previous section  one has to attach
certain weights to the various branched maps that corresponds to the
diagrams found in step (5) of section (3.4).
The weight of a  branch point that makes a
permutation of the sheets which has $k_i$ cycles of length $i\ge 2$,
and has $k_1$ marked  points is
$\lambda A$ times the sum of the coefficients of all the diagrams that
have $k_i$ loops of length $i$ (after performing
the manipulations of section 3.4), times
a symmetry factor  $\prod_{i=1}^\infty (i^{k_i}k_i!)$.
\begin{equation}
(\lambda A) \times \left(
\begin{array}{c} \mbox{Sum of coefficients of diagrams}\\
                 \mbox{with $k_i$ loops of length $i$}
\end{array}
\right)
\times
\prod_{i=1}^\infty (i^{k_i}k_i!)
\end{equation}

Returning to equation (\ref{RHOHP}), in order
to keep the overall factor of $N^{2-2h}$ which is needed for a stringy
interpretation as in (\ref{FULLGT},\ref{Hur}),
the factors $h_\rho^{\rho'}$ should be
$$
{\cal S}(\rho)+{\cal N}(\rho) -2 - {\cal S}(\rho')+{\cal N}(\rho')
$$
where we denote by ${\cal S}(\rho)$ the sum of all the summands
of the partition $\rho$ and by ${\cal N}(\rho)$
the number of summands in $\rho$.
As $\rho$ corresponds to a certain monomial in
 $\Phi(B)$, ${\cal S}(\rho)+{\cal N}(\rho)$
is the degree in $B$ of that monomial plus
the total number of traces in that monomial. Moreover, we take the
corresponding coefficient in $\Phi(B)$  to scale  as
$$
N^{2-{\cal S}(\rho)-{\cal N}(\rho)}
$$
(e.g.
$\alpha_1 N^{-1} tr(B^2) +\alpha_2 N^{-2} tr(B^3)
+ \alpha_3 N^{-4} (tr(B^2)^2)\cdots$).
In appendix (A) we will motivate this scaling also by considering the
't-Hooft like large $N$ limit in terms of Feynman diagrams for both generalized
$YM_4$ and generalized $QCD_2$.
${\cal S}(\rho')-{\cal N}(\rho')$ is the change in the Euler characteristic
of the world sheet due to a generalized branch point. A $k$-cycle branch
point adds $k-1$ to $2-2h$.
Putting all these $N$ factors together and using the Riemann-Hurwitz formula
we recover the stringy $N^{2-2h}$ behaviour. As a matter of fact,
in general we will obtain
$$
{\cal S}(\rho)+{\cal N}(\rho) -2 - {\cal S}(\rho')+{\cal N}(\rho')
 \geq h_\rho^{\rho'}
$$
 but we will not always have
equality. In order to retain the string interpretation,
this can be explained as an attachment of
$$
\hlf ({\cal S}(\rho)+{\cal N}(\rho) -2 - {\cal S}(\rho')+{\cal N}(\rho')
 - h_\rho^{\rho'})
$$
 microscopic handles to the singular points.
In the next subsection we will argue that the factors of $N$ are consistent
with attaching a microscopic Riemann surface  with at least ${\cal N}(\rho')$
handles -- each one is connected to one of the ${\cal N}(\rho')$ branch points
that correspond to $\hP_{\rho'}$. This, as we will argue, makes the string
theory local.


\subsection{Examples}
In this section we give a few examples for various choices of $\Phi(B)$ for
both $U(N)$ and $SU(N)$ groups.
\begin{enumerate}
\item
For ${\lambda\over N} tr(B^2)$ which is the conventional $YM_2$ theory we get
$$
{{2\lambda}\over N} \hP_{\{2\}} + \lambda \hP_{\{1\}}
$$
The first term means that we give a factor of ${{2\lambda A}\over N}$
for each branch point, and the second term means that we have a factor
of $\lambda$ for each marked point (i.e. this is the string tension).

\item
For $\alpha N^{-2} tr(B^3)$ in $U(N)$ we get
\begin{equation}
3\alpha N^{-2}\hP_{\{3\}} + 3\alpha N^{-1}\hP_{\{2\}}
+ 3\alpha N^{-2}\hP_{\{1+1\}} + \hlf\alpha \hP_{\{1\}}
+\hlf\alpha N^{-2}\hP_{\{1\}}
\label{EXMPB3}
\end{equation}
The first term is the contribution from branch points of degree 2
(the simple branch points are of degree 1). The next term is a modification
to the weight of the usual branch points. The third is the weight of two
marked points at the same point (but different sheets), which will translate
into $n^+ (n^+ -1)+n^- (n^- -1)$ in the weight of a map for which
$(n^+,n^-)$ are the numbers of sheets of each orientability.
The last two terms are modifications to the cosmological
constant (or, in our terminology, to the weight of the single marked point).

Note that, because of the original power of
$N^{-2}$ we do not get the usual $N^{2-2h}$ stringy behaviour
in the partition function. We can overcome this problem by interpreting
the last term not as the usual marked point,
but as a {\em microscopic handle} that is
attached to the point (this is similar to the interpretations of contracted
handles and tubes in \cite{Mina}). By interpreting certain marked points
as actually being microscopic handles (or higher Riemann surfaces) we can
always adjust the power of $N$ to be $N^{2-2h}$.
Similarly, we should interpret the term $3\alpha N^{-2}\hP_{\{1+1\}}$
as a connecting tube.
We will come back to this interpretation toward the end of this section and
investigate it further in the next subsection.

\item
For $N^{-2} tr(B^3)$ in $SU(N)$ we obtain the following corrections
to (\ref{EXMPB3}):
\begin{eqnarray}
\lefteqn{
-{6\over {N^3}}\xP{2+1}
-{{12}\over{N^3}}\xP{2}
+{{12}\over{N^4}}\xP{1+1+1}
} \nonumber\\
&-&
({6\over{N^2}}-{{12}\over{N^4}})\xP{1+1}
-({3\over{N^2}}-{2\over{N^4}})\xP{1}
\label{EXMPSUB3}
\end{eqnarray}
These terms and the terms in the previous example (\ref{EXMPB3})
 do not mix chiralities (i.e. sheets of opposite orientations).
In the full theory (chiral and anti-chiral sectors) there is the
corresponding anti-chiral term:
\begin{eqnarray}
\lefteqn{
+{6\over {N^3}}\xP{\bar{2}+\bar{1}}
+{{12}\over{N^3}}\xP{\bar{2}}
-{{12}\over{N^4}}\xP{\bar{1}+\bar{1}+\bar{1}}
} \nonumber\\
&+&
({6\over{N^2}}-{{12}\over{N^4}})\xP{\bar{1}+\bar{1}}
+({3\over{N^2}}-{2\over{N^4}})\xP{\bar{1}}
\label{EXMPACB3}
\end{eqnarray}
For $SU(N)$ there are additional
terms that do mix chiralities. They are
\begin{equation}
-{6\over {N^3}}(\xP{\bar{2}+1}-\xP{2+\bar{1}})
+{{12}\over{N^4}}(\xP{\bar{1}+\bar{1}+1}-\xP{\bar{1}+\bar{1}+1})
\label{EXMPMCB3}
\end{equation}
The first term is the contribution of maps that have a branch point
in one orientability and a marked point in the other
(at the same target space point).
The second term is the contribution of maps with three marked
points -- two for one orientability and one for the other.

To illustrate the content of these formulae in terms of
representations, we will calculate the value of the third Casimir
${1\over 6}d_{abc} t^{(a}t^b t^{c)}$ for a totally anti-symmetric
representation of $SU(N)$ with $k$ boxes. The term $\xP{3}$
is translated into the sum of all the permutations of the $k$ indices
of a totally anti-symmetric tensor that are 3-cycles, this gives
${1\over 3}k(k-1)(k-2)$. The term $\xP{2}$ gives the sum of
all the permutations that are 2-cycles, that is $-{1\over 2}k(k-1)$
(a minus sign comes from anti-symmetry).
All in all we get
\begin{eqnarray}
\lefteqn{
C_{\{3\}}(k)\equiv {1\over 6}d_{abc} t^{(a}t^b t^{c)}
} \nonumber\\
&=& k(k-1)(k-2) -{3\over 2}k(k-1) +{3\over 2}Nk(k-1)+\hlf k
+\hlf N^2 k +{3\over N}k(k-1)(k-2)
\nonumber\\
&-& 3k(k-1)+{6\over N}k(k-1)-3k
+{2\over{N^2}}k(k-1)(k-2) + {6\over{N^2}}k(k-1) +{{2k}\over{N^2}}
\nonumber\\
&=&{k\over{2N^2}}(N+1)(N+2)(N-k)(N-2k)
\end{eqnarray}

In appendix (B) we outline a different (much more lengthy) derivation
of the stringy rules of a $tr(B^3)$ insertion, using the string-winding
 creation-operators formalism of Douglas \cite{DougTec}.

\item
For $N^{-3} tr(B^4)$ in $U(N)$ we get
$$
4N^{-3}\xP{4} + 6N^{-2}\xP{3}
+6N^{-3}\xP{2+1} +{8\over 3}N^{-2}\xP{1+1}
+({4\over 3}N^{-1}+6N^{-3})\xP{2}
+({1\over 6}+{5\over 6} N^{-2})\xP{1}
$$
The terms that have an extra microscopic handle are
$$
6N^{-3}\xP{2+1} +{8\over 3}N^{-2}\xP{1+1} +6N^{-3}\xP{2}
+{5\over 6} N^{-2}\xP{1}
$$

\item
For $N^{-4}(tr(B^2))^2$ in $U(N)$ we get
\begin{eqnarray}
\lefteqn{
24N^{-4}\xP{3} + 8N^{-4}\xP{2+2} +4N^{-3}\xP{2+1}
} \nonumber\\
&+&
{{16}\over 3}N^{-3}\xP{2} +(2+{8\over 3}N^{-4})\xP{1+1}
+({2\over 3}N^{-2} +{1\over 3}N^{-4})\xP{1}
\end{eqnarray}
and additional terms:
$$
8N^{-4}\xP{2+\bar{2}} +4N^{-3}(\xP{\bar{2}+1}+\xP{2+\bar{1}})
+2N^{-2}\xP{1+\bar{1}}
$$
that mix the two chiral sectors.

The meaning of the term $2N^{-2}\xP{1+\bar{1}}$ is a factor of
$N^{-2}e^{-2\alpha n^+ n^- A}$, where $\alpha$ is the coefficient of the
$N^{-4}(tr(B^2))^2$ term in the action.
$n^+$ is the number of sheets of positive orientability and $n^-$ is the
number of sheets of negative orientability for a given map.
This extra term came from dividing the two 2-vertex loops that correspond to
$(tr(B^2)^2)$ into a chiral loop and an anti-chiral loop, in step (2) of
section (3.4).
In the next section we will interpret it, following \cite{GT},
 as a connecting tube.

\end{enumerate}


We end this section by returning to the interpretation of the powers of $N$
as microscopic attached surfaces. The consistency of such an interpretation
stems from Fig.6 (as a matter of fact it is easier to consider Figs.4a-b,
which give rise to Fig.6 by iterating them $r-1$ times).
It is straightforward to show that the reduction depicted in Fig.6 results
in an even power of $N$. Moreover, the reduction always produces a
non-positive power of $N$ which can be re-interpreted as a positive number
of attached handles. We recall that contributions to the $N$ power come
from branch points, closed loops with no vertices and the original
$N$-scaling of the coefficients of the various terms in $\Phi(B)$.
The branch points contribute $-\sum (i-1)k_i$ (for a partition
$k_1\cdot 1+k_2\cdot 2+\cdots$ while each closed loop in the diagram
(after the above steps have been performed) contributes one power of $N$.
The $({1\over N})$-scaling of the coefficients
in $\Phi(B)$ for a term of the form $tr(B^i)$ (the general term will be
treated in the next subsection) is equal to $i-1$ which is
the number of vertices in the diagrams associated with step (1)
of our construction (see section 3.4) minus $1$.
Each step in Figs.4a-b reduces the number of vertices by one and either
increases or decreases the total number of loops by one.
Thus, after each step the $N$-power remains the same (if the number
of loops is increased by one) or decreases by 2 (in case the number
of loops decreases by one).

\vskip 0.5cm
\parbox[c]{150mm}{
\begin{picture}(150,150)

\thicklines

\put(5,100){\circle{2}}
\put(15,100){\circle{2}}
\put(10,108){\circle{2}}

\put(6,100){\vector(1,0){8}}
\put(14,101){\vector(-1,2){3}}
\put(9,107){\vector(-1,-2){3}}

\thinlines
\put(5,100){\circle{8}}
\put(15,100){\circle{8}}
\put(10,108){\circle{8}}

\thicklines
\put(6,90){$3\xP{3}$}

\thicklines

\put(30,100){\circle{2}}
\put(40,100){\circle{2}}
\put(35,108){\circle{2}}

\put(31,100){\vector(1,0){8}}
\put(39,101){\vector(-1,2){3}}
\put(34,107){\vector(-1,-2){3}}

\thinlines
\put(35,100){\oval(20,5)}
\put(35,108){\circle{8}}

\thicklines
\put(21,90){$3\xP{1+1}+3N\xP{2}$}

\thicklines

\put(70,100){\circle{2}}
\put(80,100){\circle{2}}
\put(75,108){\circle{2}}

\put(71,100){\vector(1,0){8}}
\put(79,101){\vector(-1,2){3}}
\put(74,107){\vector(-1,-2){3}}

\thinlines
\put(75,103){\circle{15}}

\thicklines
\put(61,90){$\hlf\xP{1}+\hlf N^2\xP{1}$}

\thicklines

\put(5,58){\circle{2}}
\put(15,58){\circle{2}}

\put(10,57){\oval(10,4)[b]}
\put(10,59){\oval(10,4)[t]}

\put(15,57){\vector(0,1){1}}
\put(5,59){\vector(0,-1){1}}

\put(10,68){\circle{2}}
\put(9,66){\oval(4,4)[l]}
\put(11,66){\oval(4,4)[r]}
\put(9,64){\vector(1,0){2}}

\thinlines
\put(5,58){\circle{8}}
\put(15,58){\circle{8}}
\put(10,68){\circle{8}}

\thicklines
\put(2,45){$-{6\over N}\xP{2+1}$}

\thicklines

\put(35,58){\circle{2}}
\put(45,58){\circle{2}}

\put(40,57){\oval(10,4)[b]}
\put(40,59){\oval(10,4)[t]}

\put(45,57){\vector(0,1){1}}
\put(35,59){\vector(0,-1){1}}

\put(40,68){\circle{2}}
\put(39,66){\oval(4,4)[l]}
\put(41,66){\oval(4,4)[r]}
\put(39,64){\vector(1,0){2}}

\thinlines
\put(40,68){\circle{8}}
\put(40,58){\oval(18,9)}

\thicklines
\put(32,45){$-6\xP{1+1}$}

\thicklines

\put(65,58){\circle{2}}
\put(75,58){\circle{2}}

\put(70,57){\oval(10,4)[b]}
\put(70,59){\oval(10,4)[t]}

\put(75,57){\vector(0,1){1}}
\put(65,59){\vector(0,-1){1}}

\put(75,68){\circle{2}}
\put(74,66){\oval(4,4)[l]}
\put(76,66){\oval(4,4)[r]}
\put(74,64){\vector(1,0){2}}

\thinlines
\put(65,58){\circle{8}}
\put(75,63){\oval(8,18)}

\thicklines
\put(61,45){$-{{12}\over N}\xP{2}$}

\thicklines

\put(95,58){\circle{2}}
\put(105,58){\circle{2}}

\put(100,57){\oval(10,4)[b]}
\put(100,59){\oval(10,4)[t]}

\put(105,57){\vector(0,1){1}}
\put(95,59){\vector(0,-1){1}}

\put(100,64){\circle{2}}
\put(99,66){\oval(4,4)[l]}
\put(101,66){\oval(4,4)[r]}
\put(99,68){\vector(1,0){2}}

\thinlines
\put(100,59){\circle{15}}

\thicklines
\put(93,45){$-3\xP{1}$}

\thicklines

\put(10,27){\circle{2}}
\put(9,29){\oval(4,4)[l]}
\put(11,29){\oval(4,4)[r]}
\put(9,31){\vector(1,0){2}}

\put(15,19){\circle{2}}
\put(14,21){\oval(4,4)[l]}
\put(16,21){\oval(4,4)[r]}
\put(14,23){\vector(1,0){2}}

\put(5,19){\circle{2}}
\put(4,21){\oval(4,4)[l]}
\put(6,21){\oval(4,4)[r]}
\put(4,23){\vector(1,0){2}}

\thinlines
\put(5,19){\circle{6}}
\put(10,27){\circle{6}}
\put(15,19){\circle{6}}

\thicklines
\put(2,10){${{12}\over{N^2}}\xP{1+1+1}$}

\thicklines

\put(40,27){\circle{2}}
\put(39,29){\oval(4,4)[l]}
\put(41,29){\oval(4,4)[r]}
\put(39,31){\vector(1,0){2}}

\put(45,19){\circle{2}}
\put(44,21){\oval(4,4)[l]}
\put(46,21){\oval(4,4)[r]}
\put(44,23){\vector(1,0){2}}

\put(35,19){\circle{2}}
\put(34,21){\oval(4,4)[l]}
\put(36,21){\oval(4,4)[r]}
\put(34,23){\vector(1,0){2}}

\thinlines
\put(40,27){\circle{6}}
\put(40,19){\oval(20,5)}

\thicklines
\put(32,10){${{12}\over{N^2}}\xP{1+1}$}

\thicklines

\put(70,27){\circle{2}}
\put(69,29){\oval(4,4)[l]}
\put(71,29){\oval(4,4)[r]}
\put(69,31){\vector(1,0){2}}

\put(75,23){\circle{2}}
\put(74,21){\oval(4,4)[l]}
\put(76,21){\oval(4,4)[r]}
\put(74,19){\vector(1,0){2}}

\put(65,23){\circle{2}}
\put(64,21){\oval(4,4)[l]}
\put(66,21){\oval(4,4)[r]}
\put(64,19){\vector(1,0){2}}

\thinlines
\put(70,23){\circle{15}}

\thicklines
\put(64,10){${2\over{N^2}}\xP{1}$}

\end{picture}
\\
{\it Fig. 8: Application of the rules for $tr(B^3)$.
              The first row is the $U(N)$ contribution and the
              second and third rows are $SU(N)$ corrections.
              Diagrams that mix chiral and anti-chiral sectors have
              not been included.} \\ \\ } \\

\subsection{Marked points and collapsed surfaces}
Apart from the higher ordered branch points, it followed from sections 3.5 and
 3.6 that the  generalized stringy \YM theory admits
another new feature -- the {\em marked points}.

We can think of the $e^{-\hlf n \lambda A}$ factor
as arising from the contribution of maps with
marked points. A map with $k$ marked points is weighted with
the factor of ${{(\hlf \lambda A)^k}\over {k!}}$ and the $n^k$ factor
comes from the fact that the marked point can be at any one of the $n$ sheets.
For the ordinary $YM_2$ theory, there were only maps with two
marked points at the
same target space point (but different sheets) for the group $SU(N)$
and there were no  marked points for $U(N)$ \cite{Mina}. In any case, maps
with more than two marked points had weight zero.
The maps with two marked points were interpreted in \cite{Mina,GT} as maps
with microscopic connecting tubes.
It turns out that in the generalized theory, maps with multiple marked points
 at a single target space point do have a nonzero weight.
In what follows we show that they can be thought of as being the
connecting points of microscopic surfaces of genus greater than one, thus
generalizing the connecting tubes of \cite{Mina,GT}.
Note that, for \gYM, microscopic surfaces appear both for the $U(N)$ case
as well as for the $SU(N)$ case.

The interpretation we gave in the previous section to the term $\xP{\rho}$
leads to a problem with locality for the case in which $\rho$ has
more than one cycle.
For example, a term $\xP{a+b}$, corresponds to  a point in target space
in which there are two branch points, one of degree $(a-1)$ and the other of
degree $(b-1)$.
This interpretation is local in the target space,
but, as it stands, is {\em non-local} in the world-sheet,
and therefore cannot be interpreted as a {\em contact term}.
Moreover, among the geometrical objects which we introduced to account
for the large $N$ expansion, there is one object that does not
have the interpretation of a local covering map i.e. the {\em marked point}.

In this section we will show that if the coefficient of the
term of the form $\prod_i (tr(B^i))^{k_i}$ in $\Phi(B)$
scales as $N^{1-\sum k_i (i-1)}$ then
we can re-interpret the marked points as microscopic connecting
Riemann surfaces.
As a matter of fact we will show that the power of ${1\over{N^2}}$
which stands in front of a general term $\xP{\rho}$ is always the one needed
to interpret it as a connecting microscopic Riemann surface that connects
all the seemingly ``world-sheet disconnected'' branch-points.
This generalizes the notion of a microscopic connecting tube which
was first introduced in \cite{Mina} to explain the extra $SU(N)$ terms
(in comparison to the $U(N)$ case that did not have them).

In  appendix (A)  we will demonstrate that the ${1\over{N^2}}$ powers that
we associate with the various  monomials in $\Phi(B)$ is precisely the kind of
scaling that one would adopt in the large $N$ 't-Hooft like analysis
of (generalized) $YM_4$ theory. In such an analysis we expect planar diagrams
to give the leading $O(N^2)$ contribution.
The same scaling behaviour is also the appropriate one for the 't-Hooft
like large $N$ analysis of generalized $QCD_2$. As we show in appendix (A),
this scaling behaviour guarantees that planar diagrams with the topology
of just one fermion loop are leading at large $N$.

Returning to the microscopic Riemann surfaces, we will demonstrate the
interpretation for the case of
$$
\Phi(B)={1\over{N^{n-1}}}tr(B^n),
$$
for simplicity.
The general case can be worked out in the same way.
Suppose that after the manipulations of Figs.6-7,
we end up with a partition
$$
\rho = \sum_{i=1} k_i\cdot i \equiv
\underbrace{1+\cdots+1}_{k_1}
+\underbrace{2+\cdots+2}_{k_2} +\cdots
$$
and $k_0$ closed loops with no vertices, that, according to our rules,
 contribute a factor of $N^{k_0}$.
Then it can easily be seen from Figs.6-7 (and even more easily from
Figs.4a-b ) that
\begin{equation}
\sum_{i=0}^\infty k_i (i+1) \leq n+1
\label{INEQK}
\end{equation}
(Indeed, when we split a loop of $r+r'+1$ vertices into two loops
of $r$ vertices and of $r'$ vertices as in Fig.4a the LHS of (\ref{INEQK})
stays the same, whereas the step depicted in Fig.4b only
decreases the LHS of (\ref{INEQK}) by two.)
The  factor of $N$ that goes with $\rho$ is $N^{k_0 +1 -n}$
and we need $N^{-\sum_{i=1} k_i (i-1)}$ for the branch points.
Altogether, we are left with a power of
\begin{equation}
k_0+\sum_{i=1} k_i (i-1) +1 -n \le 2 - 2\sum_{i=1}^\infty k_i
\label{INEQKK}
\end{equation}
for the marked points.
If there is just one marked point, with no power of $N$ attached
to it, we interpret the coefficient of $\xP{1}$ as a modification to the
cosmological constant.
For a single marked point but with additional branch points such as
$$
\xP{1+{\underbrace{2+\cdots+2}_{k_2}} +\cdots}
$$
we see from (\ref{INEQKK}) that there will be, at least,
 a power of ${1\over{N^2}}$ attached to it,
since $\sum_{i=0}^\infty k_i \ge 2$.
Similarly, we see from (\ref{INEQKK}) that a term of the form
$$
\xP{\underbrace{1+\cdots+1}_{k_1}}
$$
will come with a factor of at least ${1\over{N^{2k_1}}}$.
Thus, it is always possible to interpret such a term, not as $k_1$ marked
points, but as a connecting Riemann surface of genus $k_1-1$ that connects
$k_1$ sheets as in Fig.9.

Moreover, our problem of non-locality in the world sheet is solved, because
a general term
$$
\xP{\underbrace{1+\cdots+1}_{k_1} + \underbrace{2+\cdots+2}_{k_2} +\cdots}
$$
has $\sum_{i=1}^\infty k_i$ different points in the world sheet
which are the same point on the target space (branch points  or marked points).
We see from (\ref{INEQKK}) that such a term always has a factor of
at least
$$
{1\over{N^{2\sum_{i=1}^\infty k_i -2}}}
$$
in front of it. This is precisely the factor that is needed for
a Riemann surface of $\sum_{i=1}^\infty k_i$ handles that connects the
$\sum_{i=1}^\infty k_i$ distinct points.
Therefore, we can interpret the term as a microscopic
collapsed Riemann surface that connects $\sum_{i=1}^\infty k_i$ points
and $\sum_{i=2}^\infty k_i$ of those points are themselves branch points
of various degrees.

\vskip 0.5cm
\parbox[c]{100mm}{
\begin{picture}(100,30)

\thicklines

\put(50,10){\oval(20,15)[t]}
\put(54,10){\oval(6,8)[t]}
\put(46,10){\oval(6,8)[t]}

\put(40,10){\line(0,-1){6}}
\put(43,10){\line(0,-1){6}}

\put(49,10){\line(0,-1){3}}
\put(51,10){\line(0,-1){3}}

\put(39,10){\line(-1,0){20}}
\put(60,10){\line(1,0){20}}
\put(57,10){\line(-1,0){5}}
\put(48,10){\line(-1,0){4}}

\put(51,7){\line(1,0){29}}
\put(49,7){\line(-1,0){5}}
\put(39,7){\line(-1,0){20}}

\put(43,4){\line(1,0){37}}
\put(40,4){\line(-1,0){20}}

\put(12,23){connecting surface}
\put(36,21){\vector(1,-1){5}}

\put(85,6){sheets}

\end{picture}
\\
{\it Fig. 9: A connecting Riemann surface of genus two.} \\ \\ }

\section{Summary and Discussion}
In this paper  we studied
 the generalized two-dimensional  Yang-Mills theory.
A generalization of the exact  formulae for
the partition function\cite{Rus,WitRev} and Wilson loop averages
\cite{Kaza,Barlic}  of the conventional YM theory were written down. These
expressions are based on a  replacement of
the second Casimir operator  with    more general Casimir
operators depending on the particular model.
Our results agree with \cite{WitRev}, where they were obtained
by a different method, i.e. by regarding the general Yang-Mills
actions as perturbations of the topological theory at zero area.
Actually, the result of \cite{WitRev} differs by a shift of $\Phi$,
due to a normal ordering ambiguity in the action.
In \cite{WitRev} the ambiguity was fixed by requiring the results
of the instanton expansion and the large area expansions to match.
We fix the ambiguity differently by a  lattice regularization  scheme.

Using the relations between $SU(N)$ representations and
representations of the symmetric groups $S_n$, we obtained
the generalizations that have to be made in the Gross-Taylor string
rules for 2D Yang-Mills theory, so as to make the generalized
Yang-Mills theory for $SU(N)$ or $U(N)$ a local string theory as well.
The extra terms are special weights for certain maps with branch points
of a degree higher than one.

An obvious  extension of the present work is to consider other gauge
groups.  The conventional \YM theory with gauge groups $O(N)$ or $Sp(N)$
was shown\cite{NRS,Ramg} to be related to
maps from non-orientable world-sheets.  A natural conjecture is that the
generalized gauge theories are associated with  higher order branched maps from
those world sheets.  Another important topic for further exploration is
the issue of the   phase transition distinguishing between the small and large
 areas behaviour, analogous to the one   discovered recently in
\cite{DugKaz,Bou,MinPol3}.
The coupling of the \gYM theories to fermionic matter was analyzed
in \cite{DougLi} in the framework of `t Hooft's analysis. This domain of
research is far from being fully explored. A particular  interesting
question is to find out certain  $\Phi(B)$-s that lead to a special behaviour
of the coupled system. This topic is under current investigation. In fact,
already in the pure gauge case one may anticipate ``non-universal" features for
particular choices.  For instance one may get
 a singularity in the partition function even without taking
the large $N$ limit.
For example, in the $U(1)$ case, the representations $R$ are labeled
by an integer $n$ and for $\Phi(B) = -\alpha\log(1 + \lambda B^2)$
 we get
$$
{\cal Z}(U(1), A) = \sum_n  (1+\lambda n^2)^{-\alpha A}
$$
which has a singularity for $2\alpha A = 1$.

Returning to the case of \gYM with
$\Phi(B)=\sum_k t_k tr(B^k)$. Clearly, upon differentiation with respect
to the coefficients $t_k$, the generalized Yang-Mills theory can be used
to calculate correlators of $tr(B^k)$ operators in $YM_2$. This is a more
efficient way to calculate such correlators than the ``straight-forward''
method described in appendix (B). The $tr(B^k)$ correlators are related
to $tr(F^k)$ correlators in $YM_2$. The precise relation is obtained
upon integration of the auxiliary $B$ field. We demonstrate it in
appendix (B) by calculating $\bra tr(B^3)\cdots\ket$ and comparing it
to a ``straight-forward'' calculation of $\bra tr(F^3)\cdots\ket$ in
the Gross-Taylor string theory. This will also demonstrate how, within the
Gross-Taylor approach, contributions from maps with higher branch points
result from maps with ``contacts'' of lower branch points.

Another interesting fact about the generalized $YM_2$ theories is that
we can probably get rid of the $\Omega$-points of \cite{GT} (which
were given the interpretation of the Euler character in \cite{CMR}) by
an appropriate choice of $\Phi(B)$ such that $\sum \lambda_\rho C_\rho(R)$
will give $(2-2G)\log (dim R)$.

One of the most challenging questions is
 to understand the different  generalized $\Phi$ actions in terms
of  string theory Lagrangians, such as the ones proposed by \cite{Horava,CMR}.
The authors of \cite{CMR} have suggested that the gravitational
descendants of the area operator are related to the higher Casimirs.
Indeed, in the algebraic-topological framework of \cite{Wtopgrv}
the $\sigma_n$-s of the topological gravity, impose certain
constraints that pick up contributions from {\em the boundary} of
the moduli space of complex structures. Those boundary terms come either
from pinched world-sheets, or from maps with branch points of higher
degrees. Thus, they are related to the $\hP_{\rho}$-s.
It is interesting to determine the exact form of the dependence
({\em under preparation}).

It seems, however, that there are much more perturbations of the form
$\Phi(B)$, for the $YM_2$ theory, than there are perturbations of the
topological sigma model coupled to topological gravity.
The $\Phi(B)$ perturbations correspond to $\hP_{\rho}$ perturbations
with an subindex $\rho$ that is an arbitrary partition, whereas the index of
the gravitational descendants is a single integer that probably
corresponds to the degree of the degeneration of the map at a certain point.
We suspect that the other perturbations (e.g. $(tr(B^2)^2)$, whose
leading string-theoretic analogue is $\xP{2+2}$) correspond to terms
that are {\em non-local} in the topological sigma model (e.g.
$\int \sigma_1^{(2)}({\cal A})
\int \sigma_1^{(2)}({\cal A})$
where ${\cal A}$ is the area operator, see \cite{CMR}).
This conjecture stems from the fact that every $\hP_{\rho}$ operator
can be written as a polynomial of the simple $\xP{n}$ operators.
For example
$$
\xP{2+2} = \hlf\xP{2}\xP{2}-\hlf\xP{3}-\hlf\xP{1}\xP{1}+\hlf\xP{1}
$$
since two different branch points that coalesce (the $\xP{2}\xP{2}$ term)
can form either two separate branch points at the same point (the $\xP{2+2}$
term) or a branch point of order 2 (the $\xP{3}$ term) or marked points
(the $\xP{1}(\xP{1}-1)$ term).
Our interpretation of the general $\hP_{\rho}$-s was in terms of collapsed
microscopic surfaces, which are too at the boundary  of the moduli space of
complex structures and holomorphic maps.
The gravitational descendant $\sigma_n({\cal A})$ should produce a
linear combination of all the possible degenerations of order $n$.
(For $n=1$ we get a combination of branch points and connecting tubes.)
So, only one combination of the $\hP_{\rho}$-s corresponds to the
single local operator $\sigma_n({\cal A})$ and the rest are probably
non-local.
It is interesting to check whether this operator corresponds
to $tr(B^{n+1})$ and operators with more traces are non-local.

If indeed \gYM is a topological theory associated with holomorphic maps
along the line of \cite{CMR}, we expect to find for this theory those
features which are common to all topological sigma models coupled to
topological gravity. In particular we should find the
integrable structure \cite{IntStr}, recursion relations\cite{RecRel},
constraint relations \cite{ConRel} and some contact algebra \cite{ConAlg}.
It would be interesting to check whether the stringy interpretation
advocated in this paper is consistent with such recursion relations.
Work in this direction is under progress.


\vspace{1.5cm}
\centerline{\bf Acknowledgment}
 We would like to thank  D. Kutasov, B. Rusakov  and A. B. Zamolodchikov
 for useful conversations.

\newpage
\section*{Appendix A: The $O({1\over N})$
   scalings of the coefficients in $\Phi(B)$}
We have seen in section (3.2) that the correct ${1\over N}$
scaling of the coefficient
of a term $\prod_i tr(B^i)^{k_i}$  in $\Phi(B)$ should be
$$
{1\over {N^{\sum k_i (i+1)-1}}}
$$
This is the only choice that will give a stringy interpretation
whose coupling constant is ${1\over N}$.
The purpose of this appendix is to show
that the same scaling  also from the 't-Hooft like large $N$ analysis
in terms of Feynmann diagrams.
We simply follow the arguments of \cite{tHooft}.
It is convenient to rescale the $B$ field to $B = N E$.

We will start with generalized $YM_2$ four dimensions,
since in two dimensions there are only global degrees of freedom.
The Lagrangian, after rescaling the $B$ field to $B = N E$, is
$$
{\cal L} = N tr(E^{\mu\nu} F_{\mu\nu}) + N^\alpha\Phi(E)
$$
$E^{\mu\nu}$ is an anti-symmetric auxiliary field and $\Phi(E)$
is just one representative term in the general expression given in
(\ref{GEN}) and which in colour-space has the form $\prod_i tr(E^i)^{k_i}$.
The function $\Phi(E)$ is a gauge invariant space-time scalar, built
out of the $E^{\mu\nu}$-s. It's precise space-time form is not relevant
for the following analysis (see \cite{DougLi}).
When we build Feynman diagrams we have propagators of the form
$\langle E A \rangle$. We will have 3-vertices of the form
$\langle E A A \rangle$ and ($\sum_i ik_i$)-vertices of the form
$\langle E E \cdots E\rangle$.
According to our rescaled Lagrangian, the $\bra EAA\ket$ vertices each carry
a factor of $N$. Using the usual double-line notation for fields in the
adjoint representation, each closed loop gives rise to a factor of $N$.
A straightforward analysis of Feynman diagrams reveals that in order to
reproduce Euler's formula:
$$
2 -2g = \mbox{Vertices} + \mbox{Faces} - \mbox{Edges}
$$
we should take for the $\Phi(E)$ vertex in the Lagrangian
\begin{equation}
\alpha = 2 - \sum_i k_i = 2 - \#\{\mbox{traces in the expression
       for the vertex}\}
\end{equation}
As we will see shortly, this choice guarantees that the leading $N$ behaviour
will be of order $O(N^2)$, coming from planar diagrams.
The $N$-dependence analysis is particularly simple for the case
$\Phi(E)=tr(E^i)$ i.e. $\sum_i k_i = 1$ and, therefore, $\alpha=1$.
By drawing some simple diagrams it is easy to be convinced that Euler's formula
holds and the leading $O(N^2)$ contributions come from all the planar diagrams.
When we have a more complicated vertex, with several traces, the situation
is more involved. In this case, a Feynman diagram which is connected in
space-time may turn out to have contributions which are disconnected in colour
space. The colour disconnected contributions have relatively more closed loops
and, therefore, more $N$ powers. Graphically, a given space time vertex of
the form $\prod_i(tr (E^i))^{k_i}$ splits into $\sum_i k_i$ distinguished
``colour-vertices'' all acting at the same space time point. An example is
depicted in Fig.A1.
\vskip 0.5cm
\putfig{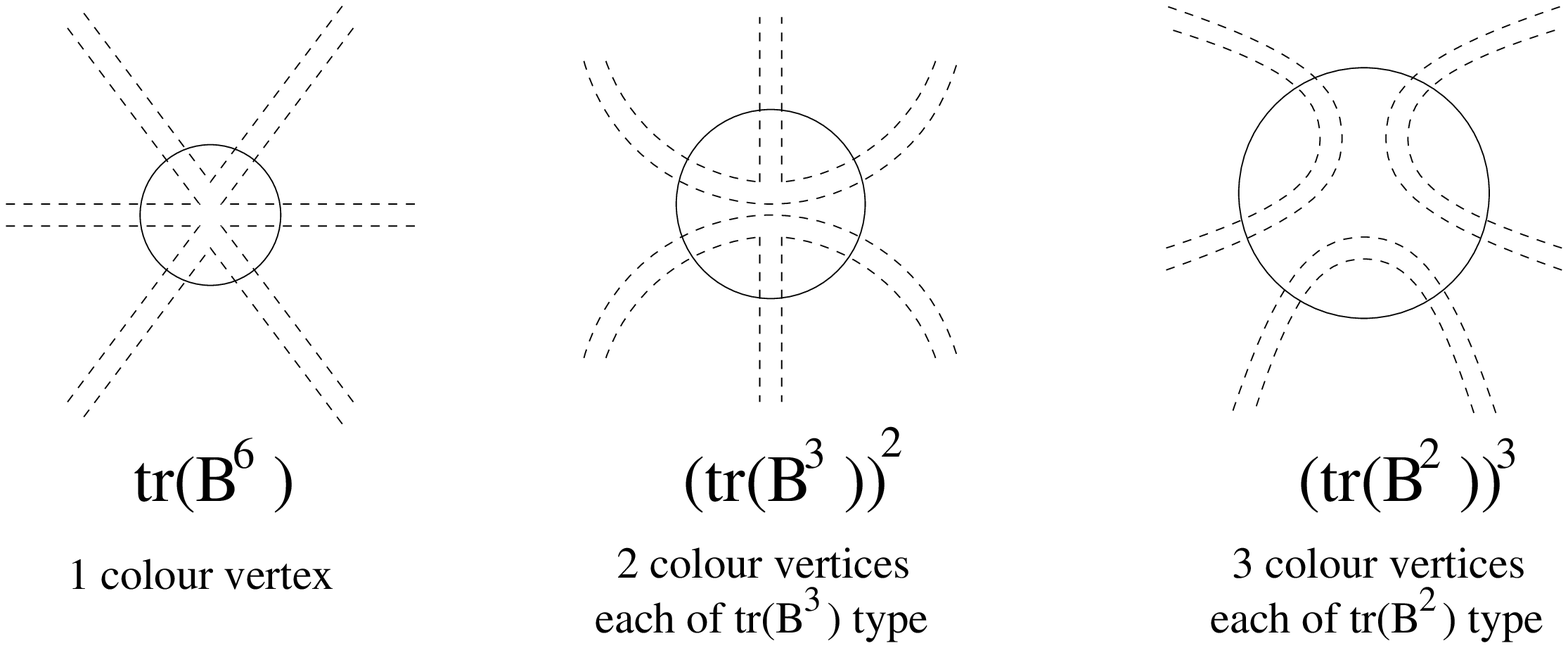}{\it Fig. A1:
           Examples of various ``colour-vertices''.}{100mm}

Let us consider a connected space-time diagram in which $q$ vertices
associated with $\Phi(E)$ appear. It is easy to see that in colour space
this diagram can split into at most $q\sum k_i -(q-1)$
distinguished disconnected colour diagrams.
If each disconnected colour part is a planar diagram and if one power of $N$
would be associated with each colour-vertex, then, just as in the simple
one trace case, we would get a contribution of $2$ from each of the
disconnected planar colour diagrams to the Euler character.
The total contribution would be:
$$
2 (q\sum k_i -(q-1)).
$$
We have assigned a power $\alpha$ to the $N$ behaviour of $\Phi(E)$.
Thus, the $N$ power would be instead
$$
2 (q\sum k_i -(q-1)) + q(\alpha - \sum k_i).
$$
(The second term accounts for the fact that each $\Phi(E)$ vertex contributes
$\alpha$ and not $\sum_i k_i$).
We choose $\alpha$ by the requirement that the maximal total $N$ power would
not exceed $2$, in accordance with Euler's formula.
{}From the last equation we read that this constraint leads to
\begin{equation}
\alpha = 2 - \sum k_i
\label{ALPHA}
\end{equation}
This is the value of $\alpha$ which we have used in section (4)
and which guarantees the stringy interpretation.

The same $\alpha$ scaling, associated with the $\Phi(E)$ vertex,
arises also in the 't-Hooft like analysis in two dimensions.
Since in 2D there are no propagating gauge degrees of freedom, we shall
consider correlators of Wilson loops. Wilson are presented
diagrammatically as closed  solid quark lines. Rather than consider the
general $\Phi(E)$ term, we'll take as an example
$\Phi(E)\sim N^\alpha tr(E^4)$. In Fig.A2 we drew a Feynman graph
contributing to the expectation value of a single Wilson loop.
\vskip 0.5cm
\putfig{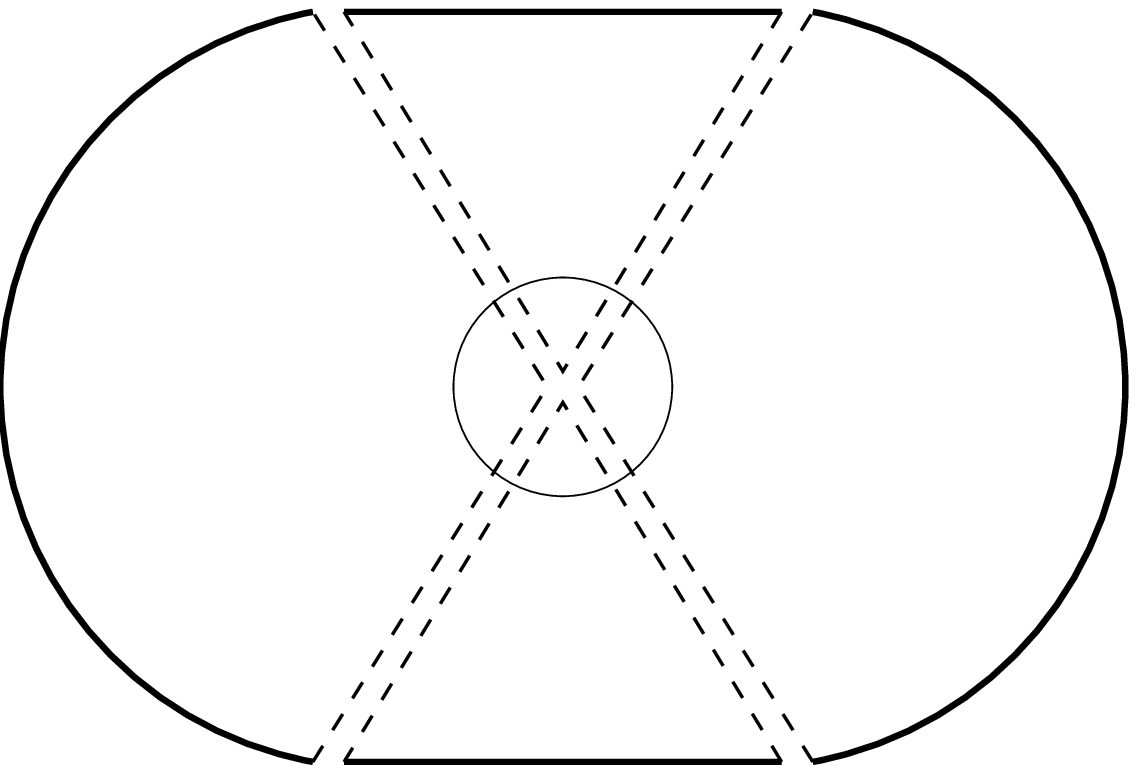}{\it
           Fig. A2: A Feynman diagram for the expectation value
           of a Wilson loop for a \mbox{$tr(B^4)$} interaction
           using the double-line notation.}{80mm}

For a stringy behaviour we want this diagram to behave as $O(N)$. This
implies $\alpha=1$, i.e. $\Phi(E)\sim N tr(E^4)$ (or equivalently
$\Phi(B)\sim N^{-3} tr(B^4)$. For $\alpha=1$ the leading $O(N)$
contributions are associated with all planar diagrams.
(For $\alpha>1$ it is straightforward to  check that as we insert more
and more $tr(E^4)$ vertices into the connected Wilson loop diagram
the $N$ power will grow. For $\alpha<1$ we will not reproduce the $O(N)$
behaviour expected of planar diagrams).

Next, we consider a $(tr(B^2))^2$ vertex. As we noted in the 4D case,
space-time terms which are made of several colour-vertices (i.e. several
traces) produce the largest $N$ power for diagrams which are connected
in space-time but disconnected in colour-space. In our example, this
situation arises for the expectation value of two Wilson loops.
(For one Wilson loop the contribution is still colour-connected
as depicted in Fig.A3).
\vskip 0.5cm
\putfig{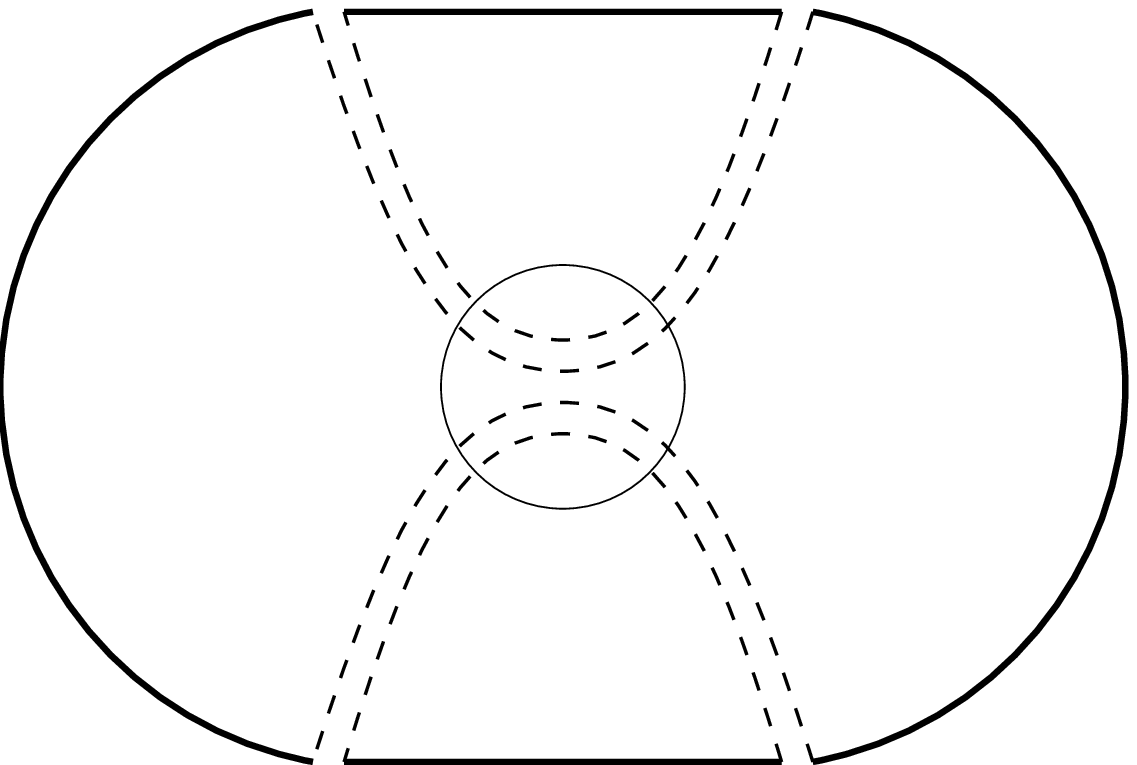}{\it
           Fig. A3: A Feynman diagram for the expetation value
           of a Wilson loop for a \mbox{$(tr(B^2))^2$} interaction
           using the double-line notation.}{80mm}

We take two Wilson loops $W_1,W_2$ which are disconnected.
We wish to calculate the connected part
of their correlator, i.e.
$$
\langle W_1 W_2 \rangle - \langle W_1 \rangle \langle W_2 \rangle
$$
For a stringy behaviour this should behave as $N^0$.
The leading contribution comes from diagrams which are colour disconnected
as in Fig.A4:
\vskip 0.5cm
\putfig{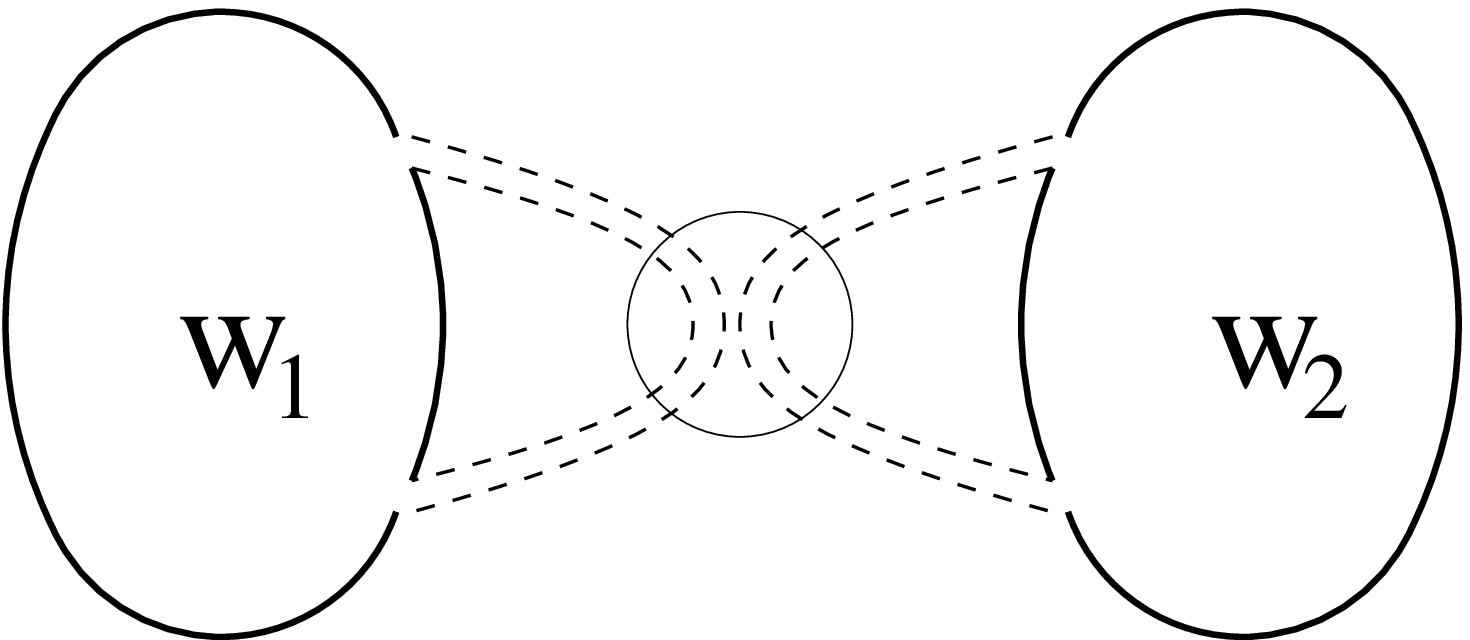}{\it
Fig. A4: A Feynman diagram for the connected
              part of the expectation value
              of the product of two Wilson loops
              for a \mbox{$(tr(B^2))^2$}
               interaction using the double-line notation.}{80mm}

It is easily seen that $\alpha=0$, i.e. $\Phi(E)\sim (tr E)^2$
(or equivalently, $\Phi(B)\sim {1\over{N^4}} (tr (B^2))^2$) produces
the desired behaviour. We can confirm the general result
$\alpha=2-\sum_i k_i$ (\ref{ALPHA}) for a vertex of the form
$N^\alpha\prod_i (tr E^i)^{k_i}$ by considering
a connected part of the expectation value of $\sum k_i$ Wilson loops.

However, after this assignment of $\alpha$, not all planar diagrams
with one fermion loop, will scale as $O(N)$. For example,
 the diagram of Fig.A3 scales as $O({1\over N})$ instead of $O(N)$.
In order to restore the usual rule that the $N$-power of a diagram
scales as the Euler characteristics of the (double-line) graph,
we will postulate that a vertex of $N^{2-\sum k_i}\prod_i (tr(E^i)^{k_i})$
has a connecting Riemann surface with $\sum_i k_i$ handles
that connect the $\sum_i k_i$ traces in $\prod_i (tr(E^i)^{k_i})$.
\vskip 0.5cm
\putfig{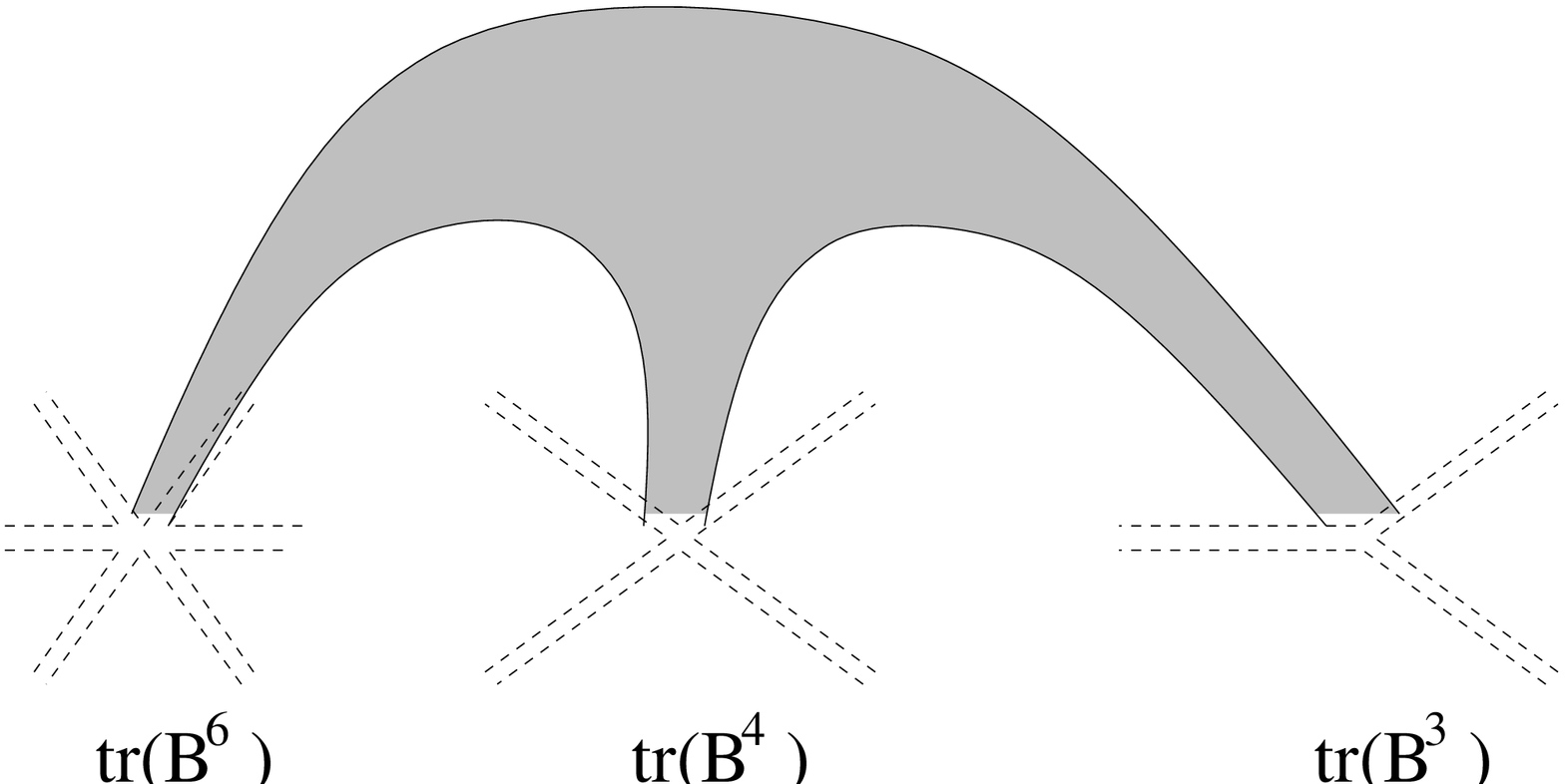}{\it Fig. A5:
We interpret a $tr(B^6)tr(B^4)tr(B^3)$ as having a connecting Riemann surface
with three handles that connects the three trace points.}{100mm}

This also has the advantage that by now there are no disconnected colour
diagrams for a diagram that is connected in space.

\section*{Appendix B}
The purpose of this appendix is to discuss the relation between the
operators $tr(B^k)$ (written in terms of the auxiliary $B$ field)
 and the operators $tr(F^k)$ in the pure YM theory.

In the body of the paper we have perturbed the $\int tr(BF) d^2x$
action by terms like $t_k \int tr(B^{k+1}) d^2x$.
where $t_2\neq 0$ and $t_k=0$ for $k>2$ corresponds to \YM.
Of course, by choosing
infinitesimal coupling constants $t_k$ we can translate the stringy rules
into rules for expectation values like:
$$
\cG(x;C_1,C_2,\dots,C_j) \equiv
\bra W(C_1) W(C_2)\cdots W(C_j) tr(B(x)^{k+1}) \ket
$$
in the {\em pure} $YM_2$ theory,
where the $W(C_i)$-s are Wilson loops and $x$ is a fixed point
which is not on either of the contours of the Wilson loops.
The stringy rules would be the same as in \cite{GT4} for the
expectation value of
$$
\bra W(C_1) W(C_2)\cdots W(C_j) \ket
$$
with new restrictions that we  sum over maps that have specified
features in $x$. For example
for $tr(B^2)$ for which we had the formula
$$
2\hP_{\{2\}} + 2\hP_{\{\bar{2}\}} + N\hP_{\{1\}} + N\hP_{\{\bar{1}\}},
$$
the maps should either have a branch point at $x$ and those are
counted with a factor of $2$, or should have a ``marked point''
(which is just a way of saying that they can be any map but
they get counted with a factor of $p(x)$ -- the total number of sheets
at $x$) and counted with a factor of $N$.

On the other hand, we can consider expressions like:
$$
\cF(x;C_1,C_2,\dots,C_j) \equiv
\bra W(C_1) W(C_2)\cdots W(C_j) tr(F(x)^{k+1}) \ket
$$
where $F(x)$ is the $YM$ field strength.
The relation between $\cG$ and $\cF$ is obtained by a simple integration.
For example:
\begin{eqnarray}
\lefteqn{
\int [DB] W(C_1) W(C_2)\cdots W(C_j) tr(B(x)^2)
  e^{\int_{\Sigma} tr(i B F - {\lambda\over N} tr(B^2)) d^2y}
=}\nonumber\\
&&
-W(C_1) W(C_2)\cdots W(C_j) ({N^2\over{4\lambda^2}}tr(F(x)^2) + N\Lambda)
e^{-{N\over {4\lambda}}\int_{\Sigma} tr(F^2) d^2y}
\end{eqnarray}
where $\Lambda$ is a cutoff, of the order of the inverse of the area of a
lattice cell (when we work on a lattice).
A similar result is obtained for the higher $tr(B^{k+1})$. The leading
term will always be
$$
({N\over{2i\lambda}})^{k+1} tr(F(x)^{k+1})
$$
and there will be corrections of lower $tr(F(x)^l)$ powers, which arise from
(a lack of) normal ordering in the $tr(F(x)^{k+1})$. We will return
to this point later on.

In this section, we will calculate the stringy rules of the $tr(F(x)^2)$
and the $tr(F(x)^3)$ insertions directly from the Gross-Taylor string
theory by considering
a tiny Wilson loop $C_x$ around $x$ in the fundamental representation,
and expanding it in the small area $\Delta$ enclosed by $C_x$.
The continuum approach gives (for a circular Wilson loop with $x$ its center):
\begin{equation}
Tr(Pe^{i\oint_{C_x} A_\alpha dx^\alpha}) \approx
1-{1\over 2}\Delta^2 Tr(F(x)^2) +{i\over 6} \Delta^3 Tr(F(x)^3)+\dots
\end{equation}
We used the fact that the matrices are in $SU(N)$ and their trace vanishes.

\subsection*{An Infinitesimal Simple Wilson Loop}
Consider the expectation value of some operators (composed of Wilson loops)
$\bra\Phi\ket$ and then the expectation value of $\bra W_x\Phi\ket$, where
$W_x$ is an infinitesimal Wilson loop surrounding the point $x$.
We can calculate this expectation value first in the string picture and then
in the $QCD_2$ picture.

We have to consider all maps with boundaries corresponding to $\Phi$ and
boundaries corresponding to $W_x$. Let $\Delta$ be the area of $W_x$ which is
infinitesimal. We will expand to order $O(\Delta^2)$.
According to \cite{GT4} we have to put one $\Omega$-point in $x$ and one
$\Omega^{-1}$-point outside $W_x$ (which we will assume is very close to
$W_x$.
The maps fall into two classes.
The first class consists of maps which are generated from the maps which
calculate $\bra\Phi\ket$ by adding a small disk $D$ whose boundary is
$W_x$. The disk can then be connected to the other sheets
by branch points and connecting tubes which may or may not be
at the $\Omega$-point. The second class is made up by maps that are generated
from maps of $\bra\Phi\ket$ by cutting out a disk $D$ from one of the sheets.

Let $n$ and $\tn$ be the operators of the cover numbers at the point $x$
for a certain map that enters in the calculation of $\bra\Phi\ket$.
For two permutations $\sigma\in S_n$ and $\tau\in S_\tn$ we define
$\bra(\sigma\otimes\tau)\Phi\ket$ to be the sum over all maps the define
$\Phi$ according to the rules of \cite{GT4} but with a slight change that
at $x$ there is a twist of $\sigma\otimes\tau$ (and {\em no} $\Omega$-point
at $x$ since we calculate just $\bra\Phi\ket$).

By the definition of \cite{GT4}, an $\Omega$-point is:
\begin{equation}
\Omega_{n\tn} = \sum_{\sigma,\tau}\sigma\otimes\tau
N^{K_\sigma+K_\tau}\prod_l V_{\sigma^{(l)},\tau^{(l)}}({l\over{N^2}})
\end{equation}
where
\begin{equation}
V_{a,b}(x)=\sum_v (-1)^v v!\binom{a}{v}\binom{b}{v}x^v
\end{equation}
$\sigma^{(l)}$ is the number of $l$-cycles in $\sigma$, and $K_\sigma$
is the total number of cycles $\sum_l \sigma^{(l)}$.
We will also use the notation of \cite{GT4}:
\begin{equation}
V_{\sigma,\tau}({1\over N})=
\prod_l V_{\sigma^{(l)},\tau^{(l)}}({l\over{N^2}})
\end{equation}
Let us also denote by $\rho_{ij}$ the permutation that switches $i$ and $j$.

We wish to expand $\bra W_x\Phi\ket$ as a power series in $\Delta$ with
coefficients of the form
$\bra C_{n,\tn}^{\sigma,\tau}\sigma\otimes\tau\Phi\ket$.

{}From \cite{GT4}(eq. 2.24) we see that we can treat connecting tubes and
contracted handles as an effective change of the factor
$e^{-{{(n+\tn)\lambda A}\over 2}}$ into
\begin{equation}
e^{-\hlf (n-{{n^2}\over{N^2}})\lambda A
-\hlf (\tn-{{\tn^2}\over{N^2}})\lambda A -{{n\tn}\over {N^2}}\lambda A}
   =e^{-\hlf (p-{{\bp^2}\over {N^2}})\lambda A}
\end{equation}
with $p=n+\tn$ and $\bp=n-\tn$

Let the $\Omega$-point be at $x$ and the $\Omega^{-1}$-point be at $y$
outside $W_x$.

The maps  of the first class
with an $\Omega$-point at $x$ and $r$ branch points inside $W_x$ and away
from $x$, and an $\Omega^{-1}$-point at $y$ contribute
\begin{eqnarray}
\lefteqn{{{(-\lambda \Delta)^r}\over{r!}}N^{1-r}e^{-\hlf\lambda \Delta}
\bra e^{{{\lambda(\bp+\hlf)\Delta}\over{N^2}}}}   \nonumber\\
&&   \sum_{\begin{array}{c}
           1\le i_1,i_2,\dots,i_r\leq n \\
                 \sigma,\tau
           \end{array}}
     \!\!\!\!\!\!\!\!\!\!\!\!\!\!\!\!\!
     (\sigma\otimes\tau)
     N^{K_{\rho_{i_1(n+1)}\cdots\rho_{i_r(n+1)}\sigma}+K_\tau-(p+1)}
     V_{\rho_{i_1(n+1)}\cdots\rho_{i_r(n+1)}\sigma,\tau}
     \Omega^{-1}(y) \Phi\ket \nonumber\\
&& \label{ADDDISK}
\end{eqnarray}
$(\sigma\otimes\tau)$ is a permutation that a ``distant~observer'' sees
on $W_x$, since for him $W_x$ looks like a single point $x$.
The factor $N^{1-r}$ is the contribution of $r$ branch points and a disk
to the genus. In the formula for $K_\sigma$ and $V$ $\tau$ is considered
as a permutation in $S_{n+1}$

The maps  of the second class contribute
\begin{equation}
e^{\hlf\lambda \Delta}\bra e^{{{\lambda(\bp+\hlf)\Delta}\over{N^2}}}
\sum_{\sigma,\tau}(\sigma\otimes\tau)N^{K_\sigma+K_\tau-p}
V_{\sigma^{(1)},\tau^{(1)}-1}({1\over{N^2}})
\prod_{l=2}^\infty V_{\sigma^{(l)},\tau^{(l)}}({l\over{N^2}})
\Omega^{-1}(y) \Phi\ket.
\label{CLASS2}
\end{equation}
However, the terms $\bra(\sigma\otimes\tau)\Omega^{-1}(y) \Phi\ket$
in (\ref{CLASS2}) include also maps with a branch point inside $W_x$
which connects the supposedly cut out sheet with some other sheet.
Such maps have to be subtracted from (\ref{CLASS2}), since we cannot cut
away a sheet which is connected to something else. But then, we would have
subtracted the contribution of maps with two branch points twice
(once for each branch point) so we have to add the contribution of maps
with two branch points, and so on. We obtain an ``inclusion-exclusion''
formula in which the contribution of $r$ branch points is:
\begin{eqnarray}
\lefteqn{{{(\lambda \Delta)^r}\over{r!}}N^{-1-r}e^{\hlf\lambda \Delta}
\bra e^{{{\lambda(\bp+\hlf)\Delta}\over{N^2}}}
          \sum_{\begin{array}{c}
           1\le j_1,j_2,\dots,j_r\leq n \\
               j_k\ne i, \tau(i)=i\\
                 \sigma,\tau
           \end{array}}
     \!\!\!\!\!\!\!\!\!\!\!\!\!\!
     \sigma\otimes\rho_{ij_1}\rho_{ij_2}\cdots\rho_{ij_r}\tau
}\nonumber\\
&\cdot&     N^{K_\sigma+(K_\tau-1)-(p-1)}
     V_{\sigma^{(1)},\tau^{(1)}-1}({1\over{N^2}})
     \prod_{l=2}^\infty V_{\sigma^{(l)},\tau^{(l)}}({l\over{N^2}})
     \Omega^{-1}(y) \Phi\ket \nonumber\\
&& \label{CUTDISK}
\end{eqnarray}
where $i$ is one of the $\tau^{(1)}$ 1-cycles of $\tau$

\subsubsection*{Formulae For $V$}
We will obtain some formulae concerning $V$.
Following \cite{DougTec}, we introduce the creation and annihilation
operators for the winding states of the string in both orientabilities.
For any function
$\psi(\sigma^{(1)},\tau^{(1)},\sigma^{(2)},\tau^{(2)},\ldots)$
we define:
\begin{eqnarray}
(\D_l^k\psi)(\sigma^{(1)},\tau^{(1)},\sigma^{(2)},\tau^{(2)},\ldots)
 &\stackrel{def}{=}&
 \psi(\sigma^{(1)},\tau^{(1)},\sigma^{(2)},\tau^{(2)},\ldots,
 \sigma^{(l-1)},\tau^{(l-1)},\sigma^{(l)}+k,\tau^{(l)},\ldots) \nonumber\\
(\bD_l^k\psi)(\sigma^{(1)},\tau^{(1)},\sigma^{(2)},\tau^{(2)},\ldots)
 &\stackrel{def}{=}&
 \psi(\sigma^{(1)},\tau^{(1)},\sigma^{(2)},\tau^{(2)},\ldots,
 \sigma^{(l)},\tau^{(l)}+k,\sigma^{(l+1)},\tau^{(l+1)},\ldots) \nonumber\\
&&
\end{eqnarray}
{}From the definition of $V$ we easily obtain the
equations:\footnote{The equations are the same as (21) of \cite{DougTec}.}
\begin{eqnarray}
l\sigma^{(l)}\D_l^{-1}V+N^2\bD_l V &=& N^2 V \label{DDB1} \\
l\tau^{(l)}\bD_l^{-1}V+N^2 \D_l V &=& N^2 V  \label{DDB2}
\end{eqnarray}
Multiplying (\ref{DDB1}) by $\D_l$ and (\ref{DDB2}) by $\bD_l$ and
subtracting we get
\begin{equation}
(l\sigma^{(l)}-l\tau^{(l)})V = N^2 (\D_l-\bD_l) V
\end{equation}
Denoting $\pbl{l}=l\sigma^{(l)}-l\tau^{(l)}$
We obtain:
\begin{eqnarray}
l\sigma^{(l)}\D_l^{-1}V &=& (N^2+\pbl{l}-N^2 \D_l)V \label{DRULES1}\\
\bD_l V &=& (\D_l + {1\over{N^2}}\pbl{l})V         \label{DRULES2}\\
l\tau^{(l)}\bD_l^{-1}V &=& N^2(1-\D_l)V            \label{DRULES3}
\end{eqnarray}
We will denote:
\begin{eqnarray}
\T_l &\stackrel{def}{=}& l\sigma^{(l)} \D_l^{-1}
\label{TL1}\\
\bT_l &\stackrel{def}{=}& l\tau^{(l)} \bD_l^{-1}
\label{TL2}
\end{eqnarray}
The multiplication of $T$ and $D$ operators is defined so that
$\D_l \T_l V = l(\sigma^{(l)}+1)V$ and $\T_l \D_l = l\sigma^{(l)}V$.
Also $\T_l^2 V = l^2\sigma^{(l)}(\sigma^{(l)}-1) \D_l^{-2} V$.

We can use the rules (\ref{DRULES1}-\ref{DRULES3}) to express complicated
multiples of the operators $\D_l$,$\bD_l$,$\T_l$,$\bT_l$ in terms of the
$\D_l$-s only (or, as will be more convenient later, in terms of the $\T_l$-s).
In this procedure we have to take care of non-commuting operators
(such as $\T_l$ and $\D_l$ or $\D_l$ and $\pbl{l}$ etc).
For example:
\begin{eqnarray*}
\T_l^2 V &=& l\sigma^{(l)}\D_l^{-1}(N^2-N^2 \D_l+\pbl{l})V   \\
        &=& N^2 l\sigma^{(l)}\D_l^{-1}V -N^2 l\sigma^{(l)}V
            +(\pbl{l}-l)l\sigma^{(l)}\D_l^{-1}V            \\
        &=& (N^2+\pbl{l}-l)(N^2+\pbl{l}-N^2 \D_l)V-N^2 l\sigma^{(l)}V
\end{eqnarray*}
We shall need two more relations which we shall now derive.
Using (\ref{TL2}) and (\ref{DDB2})
\begin{equation}
\tn V = \sum_{l=1}^\infty l\tau^{(l)}\bD_l^{-1}\bD_l V
= \sum_{l=1}^\infty \bT_l\bD_l V
= \sum_{l=1}^\infty \bT_l(1-\vNN \T_l)V
\end{equation}
and a similar equation for the barred sector. Thus
\begin{eqnarray}
\sum_{l=1}^\infty \bT_l \T_l V
                &=& N^2\sum_{l=1}^\infty\bT_l V -N^2\tn V  \label{SUMTTB1}\\
\sum_{l=1}^\infty \bT_l \T_l V
                &=& N^2\sum_{l=1}^\infty \T_l V -N^2 n V  \label{SUMTTB2}
\end{eqnarray}
Adding and subtracting we get
\begin{eqnarray}
\sum_{l=1}^\infty \bT_l \T_l V
  &=& \hlf N^2\sum_{l=1}^\infty(\T_l+\bT_l)V -\hlf N^2 p V  \label{SUMTTB3}\\
\sum_{l=1}^\infty (\T_l - \bT_l) &=& \bp \label{TMINBT}
\end{eqnarray}

\subsubsection*{The String Operators}
We wish to obtain formulae for the insertion of permutations in the string
picture. Working in the algebra of permutations  we get
\begin{eqnarray*}
\sum_{i,j}\rho_{ij}\otimes 1
&=& \sum_{i,j}(\rho_{ij}\otimes 1) \Omega(x)\Omega^{-1}(x) \\
&=& \sum_{i,j}\rho_{ij}\sigma\otimes\tau N^{K_\sigma+K_\tau}
V_{\sigma,\tau}\Omega^{-1}\\
&=& \sum_{i,j}\sigma\otimes\tau
N^{K_{\rho_{ij}\sigma}+K_\tau}V_{\rho_{ij}\sigma,\tau}\Omega^{-1}
\end{eqnarray*}
The structure of the cycles of $\sigma$ is changed by the multiplication
of $\rho_{ij}$ according to:
\begin{eqnarray}
(ij) \circ (i_1 i_2\ldots i_{l-1} i)(j_1 j_2\ldots j_{l'-1} j)
 &=& (i_1 i_2\ldots i_{l-1} i j_1 j_2\ldots j_{l'-1} j)                \\
(ij) \circ  (i_1 i_2\ldots i_{l-1} i j_1 j_2\ldots j_{l'-1} j)
 &=& (i_1 i_2\ldots i_{l-1} i)(j_1 j_2\ldots j_{l'-1} j)               \\
\end{eqnarray}
Writing symbolically
\begin{equation}
\hat{S} \stackrel{def}{=}
 \sum_{\sigma,\tau}\sigma\otimes\tau
N^{K_\sigma+K_\tau} \hat{S}V_{\sigma,\tau} \Omega^{-1}
\end{equation}
we obtain the relations
\begin{eqnarray}
\sum_{ij}\rho_{ij}\otimes 1 &=&
  \sum_{l,l'}(\vN \T_l \T_{l'} \D_{l+l'} + N \T_{l+l'} \D_l \D_{l'})
\label{RHO1}\\
\sum_{ij}1\otimes\rho_{ij} &=&
 \sum_{l,l'}(\vN\bT_l\bT_{l'}\bD_{l+l'} + N\bT_{l+l'}\bD_l\bD_{l'})
                                                               \label{RHO2}
\end{eqnarray}
Finally, using (\ref{SUMTTB3}) we obtain
\begin{eqnarray}
\sum_{ij}(\rho_{ij}\otimes 1 + 1\otimes\rho_{ij}) =
\vN\sum_{l,l'}(\T_l \T_{l'} + \bT_l\bT_{l'} - 2\bT_{l'}\T_l)
+N\sum_l l(\bT_l + \T_l) - Np \nonumber\\
\label{OPRT2}
\end{eqnarray}

\subsubsection*{The Zero  Branch Points Contribution}
According to (\ref{ADDDISK}) and (\ref{CUTDISK}) for $r=0$ we get
\begin{eqnarray}
\lefteqn{
Ne^{-\hlf\lambda \Delta}
\bra e^{{{\lambda(\bp+\hlf)\Delta}\over{N^2}}}
\sum_{\sigma,\tau}(\sigma\otimes\tau)N^{K_\sigma+K_\tau-p}
\D_1 V \Omega^{-1}(y) \Phi\ket
}\nonumber\\
&&+\vN e^{\hlf\lambda \Delta}
\bra e^{{{\lambda(\bp+\hlf)\Delta}\over{N^2}}}
\sum_{\sigma,\tau}(\sigma\otimes\tau)N^{K_\sigma+K_\tau-p}
\tau^{(1)}\bD_1^{-1}V \Omega^{-1}(y) \Phi\ket
\end{eqnarray}
which according to (\ref{DRULES1}-\ref{DRULES3}) gives
\begin{eqnarray}
&-&2N\sinh\hlf\lambda \Delta
\bra e^{{{\lambda(\bp+\hlf)\Delta}\over{N^2}}}
\sum_{\sigma,\tau}(\sigma\otimes\tau)N^{K_\sigma+K_\tau-p}
\D_1 V \Omega^{-1}(y) \Phi\ket \nonumber\\
&+&Ne^{\hlf\lambda \Delta}
\bra e^{{{\lambda(\bp+\hlf)\Delta}\over{N^2}}}\Phi\ket
\label{ZEROBP}
\end{eqnarray}

\subsubsection*{The One   Branch Point Contribution}
{}From (\ref{ADDDISK}) and (\ref{CUTDISK}) for $r=1$ we get
\begin{eqnarray}
&-&{{\lambda \Delta}\over{N}}e^{-\hlf\lambda \Delta}
\bra e^{{{\lambda(\bp+\hlf)\Delta}\over{N^2}}}
\sum_{\sigma,\tau}\sigma\otimes\tau N^{K_\sigma+K_\tau-p}
\sum_{l=1}^\infty
l\sigma^{(l)} \D_l^{-1}\D_{l+1}V \Omega^{-1}(y) \Phi\ket \nonumber\\
&+&{{\lambda \Delta}\over{N}}e^{\hlf\lambda \Delta}
\bra e^{{{\lambda(\bp+\hlf)\Delta}\over{N^2}}}
\sum_{\sigma,\tau}\sigma\otimes\tau N^{K_\sigma+K_\tau-p}
\sum_{l=2}^\infty
l\tau^{(l)} \bD_l^{-1}\bD_{l-1}V \Omega^{-1}(y) \Phi\ket \nonumber\\
&&
\end{eqnarray}
We used the fact that $\rho_{i(n+1)}$ turns one $l$-cycle of $\sigma$
into an $(l+1)$-cycle.

Using the notation of (\ref{TL1}-\ref{TL2}) we write it as
\begin{eqnarray}
&-&{{\lambda \Delta}\over{N}}e^{-\hlf\lambda \Delta}
\bra e^{{{\lambda(\bp+\hlf)\Delta}\over{N^2}}}
\sum_{\sigma,\tau}\sigma\otimes\tau N^{K_\sigma+K_\tau-p}
\sum_{l=1}^\infty
\T_l \D_{l+1} V \Omega^{-1}(y) \Phi\ket \nonumber\\
&+&{{\lambda \Delta}\over{N}}e^{\hlf\lambda \Delta}
\bra e^{{{\lambda(\bp+\hlf)\Delta}\over{N^2}}}
\sum_{\sigma,\tau}\sigma\otimes\tau N^{K_\sigma+K_\tau-p}
\sum_{l=2}^\infty
\bT_l\bD_{l-1}V \Omega^{-1}(y) \Phi\ket \nonumber\\
&& \label{ONEBP}
\end{eqnarray}

\subsubsection*{The Two   Branch Points Contribution}
In (\ref{ADDDISK}) we have to find the cycle structure of
\begin{equation}
\rho_{i_1(n+1)}\rho_{i_2(n+1)}\sigma =
      \left\{\begin{array}{cl}
             (i_1 i_2 (n+1))\sigma & \mbox{for $i_1\ne i_2$} \\
             \sigma                & \mbox{for $i_1 = i_2$}
             \end{array}\right .
\end{equation}
(and a similar expression for (\ref{CUTDISK}))
the $3$-cycle term $(i_1 i_2 (n+1))$ can do one of two things:
\begin{itemize}
\item
turn an $l$-cycle and an $l'$-cycle into an $l+l'+1$ cycle:
\begin{equation}
(ij(n+1)) \circ (i_1 i_2\ldots i_{l-1} i)(j_1 j_2\ldots j_{l'-1} j)(n+1)
 = (i_1 i_2\ldots i_{l-1} i j_1 j_2\ldots j_{l'-1} j (n+1))
\end{equation}
\item
turn one $l+l'$ cycle into two cycles, and $(l+1)$-cycle and an $l'$-cycle:
\begin{equation}
(ij(n+1)) \circ (i_1 i_2\ldots i_{l-1} i j_1 j_2\ldots j_{l'-1} j)(n+1)
 = (i_1 i_2\ldots i_{l-1} i (n+1)) (j_1 j_2\ldots j_{l'-1} j)
\end{equation}
\end{itemize}
Thus, we obtain for two branch points the contribution of
\begin{eqnarray}
\lefteqn{
{{(\lambda \Delta)^2}\over{2!N^2}}e^{-\hlf\lambda \Delta}
\bra e^{{{\lambda(\bp+\hlf)\Delta}\over{N^2}}}
\sum_{\sigma,\tau}\sigma\otimes\tau N^{K_\sigma+K_\tau-p}
\{N n\D_1 V
}\nonumber\\
&+& N\sum_{l,l'} \T_{l+l'} \D_{l+1} \D_{l'} V
+{1\over N}\sum_{l,l'}
\T_l \T_{l'} \D_{l+l'+1}V\}\Omega^{-1}(y) \Phi\ket \nonumber\\
\lefteqn{
+{{(\lambda \Delta)^2}\over{2!N^2}}e^{\hlf\lambda \Delta}
\bra e^{{{\lambda(\bp+\hlf)\Delta}\over{N^2}}}
\sum_{\sigma,\tau}\sigma\otimes\tau N^{K_\sigma+K_\tau-p}
\{ {1\over N}(\tn-1)\bT_1 V
}\nonumber\\
&+& N\sum_{l,l'}\bT_{l+l'+1}\bD_l\bD_{l'}V
+{1\over N}
\sum_{l,l'}\bT_{l+1}\bT_{l'}\bD_{l+l'}V\}\Omega^{-1}(y) \Phi\ket \nonumber\\
&&\label{TWOBP}
\end{eqnarray}

\subsubsection*{The Three Branch Points Contribution}

Similarly, we obtain for three branch points the contribution of
\begin{eqnarray}
\lefteqn{
-{{(\lambda \Delta)^3}\over{3!N^3}}e^{-\hlf\lambda \Delta}
\bra e^{{{\lambda(\bp+\hlf)\Delta}\over{N^2}}}
\sum_{\sigma,\tau}\sigma\otimes\tau N^{K_\sigma+K_\tau-p}
\{
{1\over{N^2}} \sum_{l,l',l''}\T_l \T_{l'} \T_{l''}\D_{l+l'+l''+1}V
}     \nonumber\\
&+&\sum_{l,l',l''}\T_{l''}\T_{l+l'}\D_{l'}\D_{l+l''+1}V
+\sum_{l,l',l''}\T_{l}\T_{l'+l''}\D_{l''}\D_{l+l'+1}V
+\sum_{l,l',l''}\T_{l'}\T_{l+l''}\D_{l+1}\D_{l'+l''}V
\nonumber\\
&+&\sum_{l,l',l''}\T_{l+l'+l''}\D_{l+l'+l''+1}V
+N^2\sum_{l,l',l''}\T_{l+l'+l''}\D_{l+1}\D_{l'}\D_{l''}V
+\sum_{l,l'} \T_l \T_{l'} \D_{l+l'} \D_1 V
\nonumber\\
&+& N^2\sum_{l,l'} \T_{l+l'} \D_l \D_{l'} \D_1 V
+(2n-1)\sum_l \T_l \D_{l+1} V
\}\Omega^{-1}(y) \Phi\ket
\nonumber\\
+\lefteqn{
{{(\lambda \Delta)^3}\over{3!N^3}}e^{\hlf\lambda \Delta}
\bra e^{{{\lambda(\bp+\hlf)\Delta}\over{N^2}}}
\sum_{\sigma,\tau}\sigma\otimes\tau N^{K_\sigma+K_\tau-p}
\{
N^2\sum_{l,l',l''}\bT_{l+l'+l''+1}\bD_l\bD_{l'}\bD_{l''}V
} \nonumber\\
&+&\sum_{l,l',l''}\bT_{l'}\bT_{l+l''+1}\bD_{l''}\bD_{l+l'}V
+\sum_{l,l',l''}\bT_{l''}\bT_{l+l'+1}\bD_{l}\bD_{l'+l''}V
+\sum_{l,l',l''}\bT_{l+1}\bT_{l'+l''}\bD_{l'}\bD_{l+l''}V
\nonumber\\
&+&\sum_{l,l',l''}\bT_{l+l'+l''+1}\bD_{l+l'+l''}V
+{1\over{N^2}} \sum_{l,l',l''}\bT_{l+1}\bT_{l'}\bT_{l''}\bD_{l+l'+l''}V
+N\sum_{l,l'} \bT_1\bT_{l+l'}\bD_l\bD_{l'}V
\nonumber\\
&+&\vNN\sum_{l,l'}\bT_1\bT_l\bT_{l'}\bD_{l+l'}V
+(2\tn-3)\sum_l \bT_{l+1}\bD_l V \}\Omega^{-1}(y) \Phi\ket
\nonumber\\
&&\label{ThREEBP}
\end{eqnarray}

We will calculate the contributions up to $0(\Delta^3)$
\subsubsection*{The $0(\Delta)$ Order}
It is convenient to express all the $\D_l$-s and $\bD_l$-s in the sums
in terms of $\T_l$-s and $\bT_l$-s since
$\lim_{l\rightarrow\infty} \T_l V = 0$ and likewise for $\bT_l$.
We obtain
\begin{equation}
{N\over 2}(1-\vNN)
\end{equation}
\subsubsection*{The $0(\Delta^2)$ Order}
Again it is convenient to express all the $\D_l$-s and $\bD_l$-s in the sums
in terms of $\T_l$-s and $\bT_l$-s since
$\lim_{l\rightarrow\infty} \T_l V = 0$ and likewise for $\bT_l$.
We obtain
\begin{equation}
(\lambda \Delta)^2 \bra
\sum_{\sigma,\tau}\sigma\otimes\tau N^{K_\sigma+K_\tau-p}
R_{\sigma\tau}^{(2)}\Omega^{-1}(y) \Phi\ket
\end{equation}
Using (\ref{OPRT2}) (\ref{SUMTTB2}) (\ref{TMINBT}) as well as
\begin{equation}
\bT_1 \bp = \bT_1\sum_l l(\sigma^{(l)}-\tau^{(l)})=(\bp+1)\bT_1
\end{equation}
we obtain
\begin{eqnarray}
\lefteqn{R_{\sigma\tau}^{(2)}=}                               \nonumber\\
{1\over{2N}}p-{1\over{2N^3}}\bp^2+{N\over 2}\{\hlf(1-\vNN)\}^2
&+&{1\over{2N^2}}\sum_{ij}(\rho_{ij}\otimes 1 + 1\otimes\rho_{ij})
\nonumber\\ &&
\end{eqnarray}

\subsubsection*{Comparison with the previous results}
The result we have obtained from the previous sections can be
written symbolically as
\begin{eqnarray}
\lefteqn{ Tr Pe^{i\oint_\gamma \Delta^\alpha dx_\alpha} = } \nonumber\\
N\{1+\lambda \Delta(1-\vNN) &+& \hlf(\lambda \Delta)^2\{\hlf(1-\vNN)\}^2\}
\nonumber \\
+(\lambda \Delta)^2\{
{1\over{2N}}p-{1\over{2N^3}}\bp^2
&+&{1\over{2N^2}}\sum_{ij}(\rho_{ij}\otimes 1 + 1\otimes\rho_{ij})
\}+0(\Delta^3)
\label{FORMULAX}
\end{eqnarray}
By the way, note that, in our previous notation:
\begin{eqnarray}
\xP{2} &=& \sum_{ij}(\rho_{ij}\otimes 1) \nonumber\\
\xP{\bar{2}} &=& \sum_{ij}(1\otimes\rho_{ij}) \nonumber
\end{eqnarray}

We can write (\ref{FORMULAX}) as
\begin{eqnarray}
\lefteqn{ e^{-\hlf\lambda \Delta C_2(F)}
Tr_F Pe^{i\oint_\gamma A^\alpha dx_\alpha} = } \nonumber\\
+(\lambda \Delta)^2\{
{1\over{2N}}p-{1\over{2N^3}}\bp^2
&+&{1\over{2N^2}}\sum_{ij}(\rho_{ij}\otimes 1 + 1\otimes\rho_{ij})
\}+0(\Delta^3)
\label{STRINGF2}
\end{eqnarray}
where $C_2(F)=1-\vNN$ is the second Casimir of the fundamental representation.
However, here is a subtle point.
If we expand the Wilson loop in powers of $\Delta$ we get
\begin{equation}
Tr_F Pe^{i\oint_\gamma A^\alpha dx_\alpha}
=N-{1\over 2}\Delta^2 Tr(f^2)+0(\Delta^3)
\label{INFIWIL}
\end{equation}
If we differentiate the result of \cite{GT4} for the partition function
with respect to $\lambda$ we should get the string picture interpretation
of an operator insertion of $\int Tr(f^2)d^x$ and performing the
differentiation gives
\begin{equation}
{1\over{2N}}p-{1\over{2N^3}}\bp^2
+{1\over{2N^2}}\sum_{ij}(\rho_{ij}\otimes 1 + 1\otimes\rho_{ij})
\end{equation}
The last term, when integrated over an area, gives the number of branch
points in that area.

The discrepancy between the two calculations (for example (\ref{INFIWIL})
has no $0(\Delta)$ term but (\ref{STRINGF2}) has one) can be explained
as follows:
The term that
governs the fluctuations of the holonomy $U$ in the action
of a small region of area $\Delta$ is $e^{-Tr(f^2)\Delta}$ which
means that the fluctuations go like $\delta f^2 \Delta \sim 1$ and thus
$U\approx 1+Tr(f^2)\Delta^2 \sim 1+0(\Delta)$. We see that although almost
everywhere the field approximation to $U$ ($U\approx 1+A_\mu dx^\mu$) is
valid, it may not be valid at single points for which the field strength
fluctuations is large ($0({1\over \Delta})$).
%
%
We can observe from (\ref{STRINGF2}) that a plausible renormalization
of $Tr_F Pe^{i\oint_\gamma A^\alpha dx_\alpha}$ on a loop of area $\Delta$ is
$e^{-\hlf\lambda \Delta C_2(F)} Tr_F Pe^{i\oint_\gamma A^\alpha dx_\alpha}$.
The renormalizing factor is $e^{-\hlf\lambda \Delta C_2(F)}$.

This factor is correct for all orders (and for other representations as
well) since for the string vacuum (i.e. $n=\tn=0$)
the expectation value of a small Wilson loop is
$e^{-\hlf\lambda \Delta C_2(F)}$.
(The string vacuum is projected out by taking the area of the whole
target space to infinity).

\subsubsection*{The $0(\Delta^3)$ Order}
Similarly to (\ref{RHO1}-\ref{RHO2}) we obtain
\begin{eqnarray}
\lefteqn{
\sum_{ijk}\rho_{ijk}\otimes 1 =
  \sum_{l,m,k}(
  \vNN \T_l \T_m \T_k \D_{l+m+k} + N^2 \T_{l+m+k} \D_l \D_m \D_k
}\nonumber\\
&+& \T_{l+m}\T_k \D_m \D_{l+k} + \T_l \T_{m+k} \D_k \D_{l+m}
 + \T_{l+k}\T_m \D_{m+k} \D_l + \T_{l+m+k} \D_{l+m+k})\nonumber\\
&&         \label{R3HO1}\\
\lefteqn{
\sum_{ijk}1\otimes\rho_{ijk} =
  \sum_{l,m,k}(
  \vNN\bT_l\bT_m\bT_k\bD_{l+m+k} + N^2 \bT_{l+m+k}\bD_l\bD_m\bD_k
}\nonumber\\
&+& \bT_{l+m}\bT_k\bD_m\bD_{l+k} + \bT_l\bT_{m+k}\bD_k\bD_{l+m}
 + \bT_{l+k}\bT_m\bD_{m+k}\bD_l + \bT_{l+m+k}\bD_{l+m+k})\nonumber\\
&&  \label{R3HO2}\\
\end{eqnarray}
After a tedious calculation, we get the terms of
(\ref{EXMPB3},\ref{EXMPSUB3},\ref{EXMPACB3},\ref{EXMPMCB3}).

\end{document}